# Demonstrating Electrochemical $CO_2$ Capture on Redox-Active Metal-Organic Frameworks

Iuliia Vetik,[a] Nikita Žoglo,[b] Akmal Kosimov,[c] Ritums Cepitis,[c] Veera Krasnenko,[d] Huilin Qing,[e] Priyanshu Chandra,[e] Katherine Mirica,[e] Ruben Rizo,[c] Enrique Herrero,[c] Jose Solla-Gullón,[c] Teedhat Trisukhon,[f] Jamie W. Gittins,[f] Alexander C. Forse,[f] Vitali Grozovski,[a] Nadezda Kongi[a]† and Vladislav Ivaništšev,[a,g]†

Addressing climate change calls for action to control $CO_2$ pollution. Direct air capture offers a solution to this challenge. Making carbon capture competitive with alternatives, such as forestation and mineralisation, requires fundamentally novel approaches and ideas. One such approach is electrosorption, which is currently limited by the availability of suitable electrosorbents. In this work, we introduce copper-2,3,6,7,10,11-hexahydroxytriphenylene ($Cu_3(HHTP)_2$) metal-organic framework (MOF) that can act as electrosorbent for $CO_2$ capture, thereby expanding the palette of materials that can be used for this process. $Cu_3(HHTP)_2$ is the first MOF to switch its ability to capture and release $CO_2$ in aqueous electrolytes. By using cyclic voltammetry (CV), electrochemical impedance spectroscopy (EIS), galvanostatic charge-discharge (GCD) analysis, and differential electrochemical mass spectrometry (DEMS), we demonstrate reversible $CO_2$ electrosorption. Based on density functional theory (DFT) calculations, we provide atomistic insights into the mechanism of electrosorption and conclude that efficient $CO_2$ capture is facilitated by a combination of redox-active copper atom and aromatic HHTP ligand within $Cu_3(HHTP)_2$. By showcasing the applicability of $Cu_3(HHTP)_2$ – with a $CO_2$ capacity of 2 mmol $g^{-1}$ and an adsorption enthalpy of −20 kJ $mol^{-1}$, this study encourages further exploration of conductive redox-active MOFs in the search for superior $CO_2$ electrosorbents.

## Introduction

As a result of human activity, numerous environmental problems have been created. Global warming driven by $CO_2$ emissions constitutes the most pressing problem. This escalating issue poses severe threats to sustainability, making it a critical challenge to not only capture atmospheric $CO_2$ but also to make use of it.[1–4] Direct air capture is central for reducing atmospheric $CO_2$ concentration to target pre-industrial levels.[5–10] In carbon capture, there are two typically utilised methods – absorption and adsorption.[11] On the one hand, absorption involves the dissolution of $CO_2$ into the bulk medium through chemical reactions, such as those with polyamines, or through strong physical interactions, for example, in ionic liquids.[12,13] Conversely, adsorption relies on weak physical attraction for $CO_2$ to adhere to a surface, such as in zeolites.[14,15] Recently, electrosorption has emerged as a promising alternative, combining strong chemisorption at specific applied potentials and weak physisorption at open circuit potentials.[16–23] This approach has enabled the development of high-performance $CO_2$ capture materials, such as polyanthraquinone (AQ).

In figure 1 we summarised the previous progress in $CO_2$ sorbents comparing their capture capacity and sorption-desorption energy. As seen from the graph, AQ stands at the Pareto forefront with high sorption energy and capacity, similarly to polyamines. Ionic liquids and zeolites stand at the lower edge of the same front with low sorption energy and capacity. The green region above the Pareto forefront is the global target, that corresponds to the best trade-off between the capacity and energy expenditure.[35]

Electrosorption-based $CO_2$ capture represents an economically viable approach to combating the issue of global warming – a recent assessment finds the electrochemical-based technologies to outperform others in all scenarios, especially when powered with renewable electricity.[23] However, due to the novelty of the technology, the range of known materials is limited and state-of-the-art electrosorbents have mainly been studied in non-aqueous solutions and inert atmospheres.[36-38] Moreover, in known materials, the electrosorption mechanism typically involves a single variable – a specific redox-active centre – inherently restricting the chemical space of suitable compositions.[39-42]

[a] Institute of Chemistry, University of Tartu, Tartu 50411, Estonia.
[b] RedoxNRG OÜ, Narva-Jõesuu 29021, Estonia.
[c] Institute of Electrochemistry, University of Alicante, Apdo. 99, 03080, Alicante, Spain.
[d] Institute of Physics, University of Tartu, Tartu 50411, Estonia.
[e] Department of Chemistry, Burke Laboratory, Dartmouth College, New Hampshire 03755, United States.
[f] Yusuf Hamied Department of Chemistry, University of Cambridge, Cambridge CB2 1EW, UK.
[g] Department of Chemistry, University of Latvia, Riga LV-1004, Latvia.
† Corresponding authors.

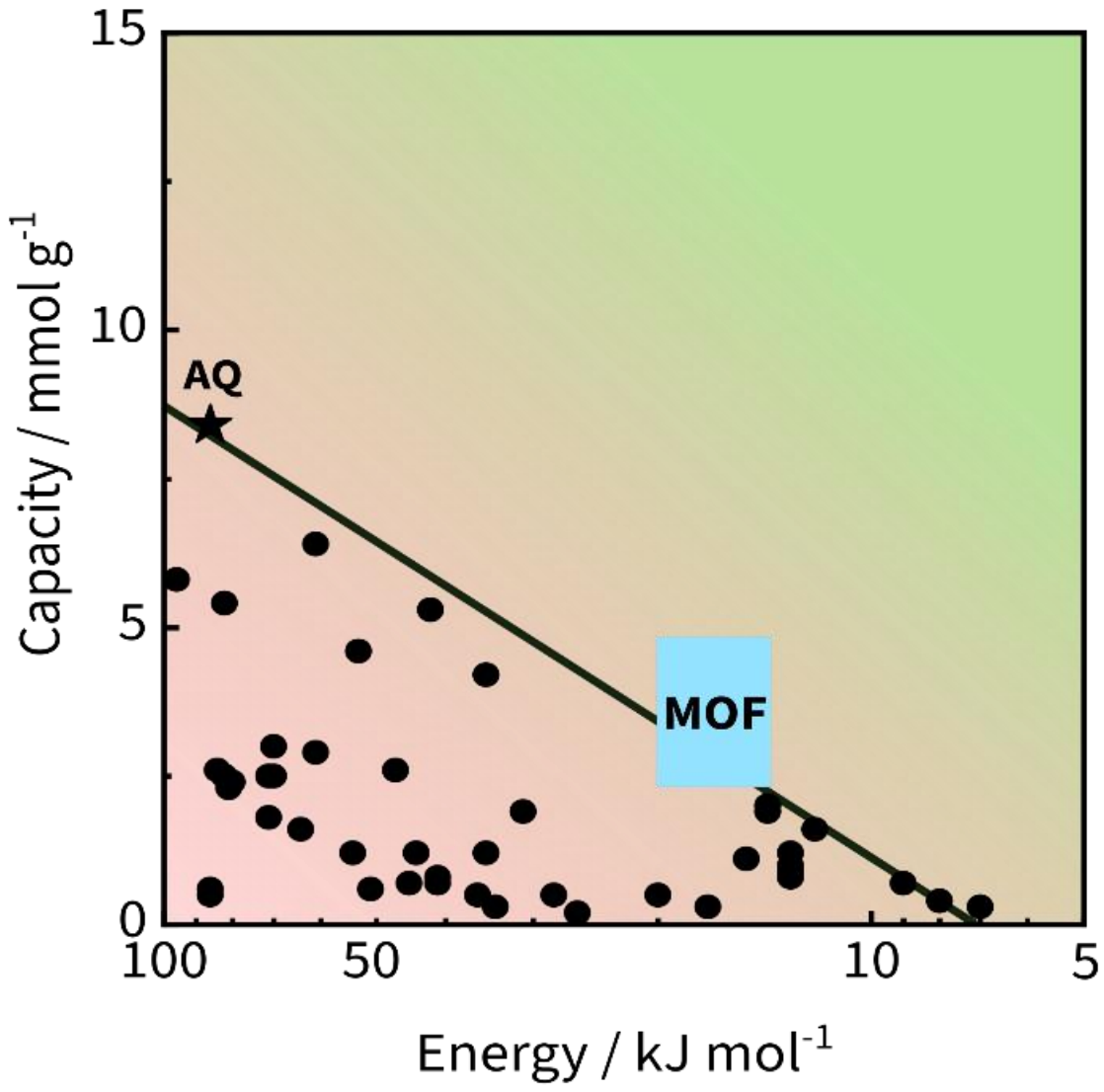

**Figure 1.** Diagram of state-of-the-art $CO_2$ sorbents in terms of sorption capacity and energy. Polyanthraquinone (AQ, star),[24] various ionic liquids, and $Cu_3(HHTP)_2$ (MOF, rectangle) are on the frontline. At the same time, other capture–release materials fall behind.[25-34] $Cu_3(HHTP)_2$ is characterised by adsorption energy of 14–20 kJ mol$^{-1}$, as obtained through DFT modelling and CV measurements, and $CO_2$ capacity of 2.4 – 4.8 mmol g$^{-1}$, as estimated from CV measurements under the assumption that the transfer of 0.5 – 1 electrons induces adsorption of one $CO_2$ molecule. $Cu_3(HHTP)_2$ is the closest material to the desired region, as indicated by the gradient.

In this context, Metal-2,3,6,7,10,11-hexahydroxytriphenylene (Metal-HHTP) metal-organic frameworks (MOFs) are increasingly recognised for their electrochemical applications, including batteries, supercapacitors, and sensors,[43-48] since they combine redox properties of metals and aromaticity of HHTP (Fig. 2a). Having two variables in redox-active MOFs – metals and ligands – squares the corresponding chemical space of possible compositions, greatly expanding the scope of applicable materials. Metal-HHTP MOFs are characterised by crystalline structure, nanoscale porosity, and extensive surface area.[49,50] Moreover, the conjugated electronic structure of the HHTP ligand facilitates both in-plane π–d interactions and out-of-plane π–π stacking, resulting in high electrical conductivity.[49-52] Such a combination of straight diffusion pathways, high electron mobility, and dense redox-active sites in conductive MOFs makes them a promising new class of tunable porous materials for $CO_2$ electrosorption. Namely, assuming that all oxygen (O) atoms within Metal-HHTP MOFs are capable of $CO_2$ electrosorption – capacities up to 14.5 mmol of $CO_2$ per g of MOF can hypothetically be achieved. Thus, we have chosen to

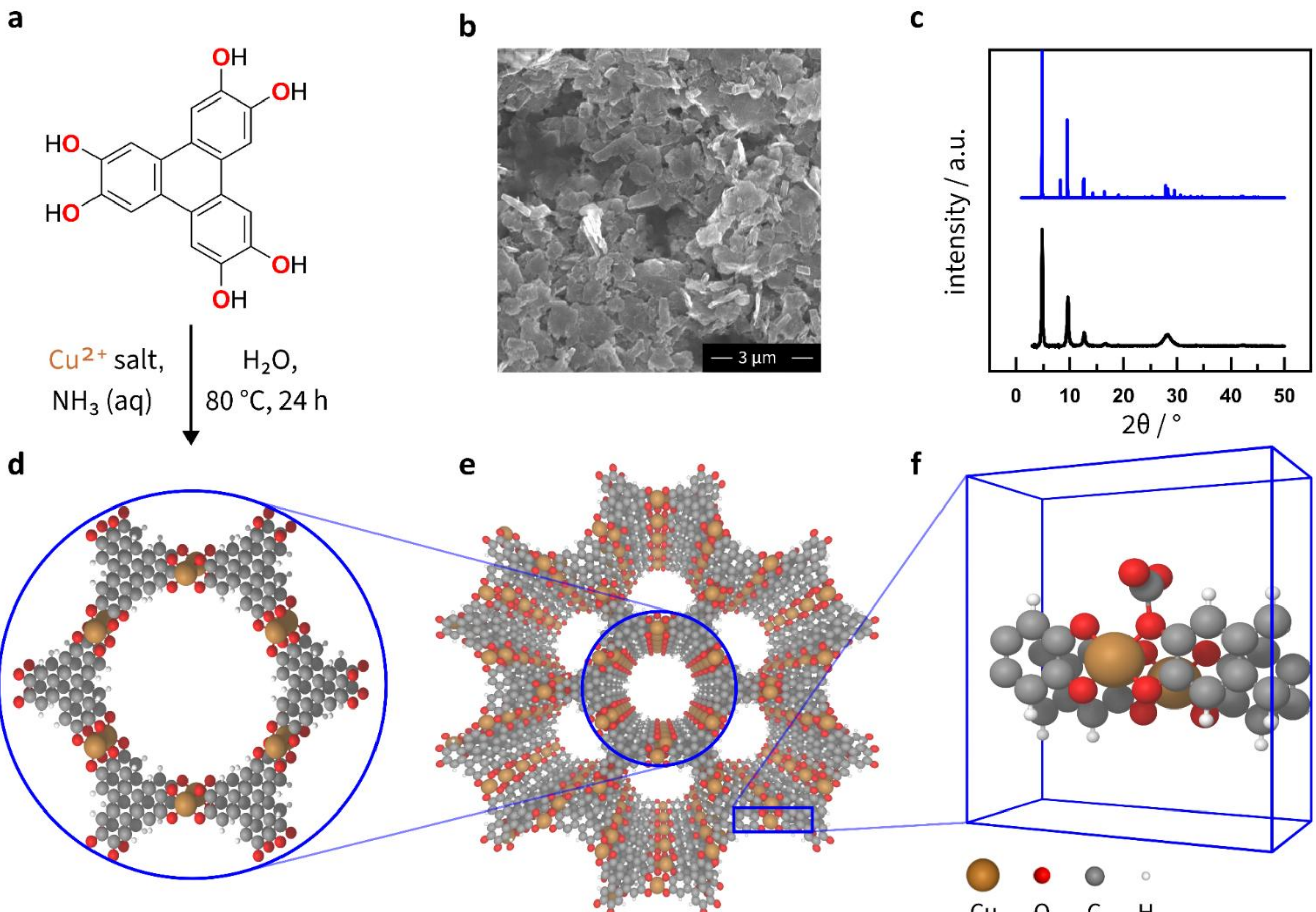

**Figure 2 (a)** Scheme for the hydrothermal synthesis of $Cu_3(HHTP)_2$, **(b)** SEM image of $Cu_3(HHTP)_2$, **(c)** simulated (blue) and experimental (black) PXRD patterns for $Cu_3(HHTP)_2$. **(d)** Schematic representation of the hexagonal structure of the $Cu_3(HHTP)_2$, **(e)** Top view of the $Cu_3(HHTP)_2$ structure with a slipped-parallel stacking mode. **(f)** A simplified model used for DFT-based simulations of $CO_2$ adsorption. Grey spheres represent C, red – O, ochre – Cu, and white – H atoms.

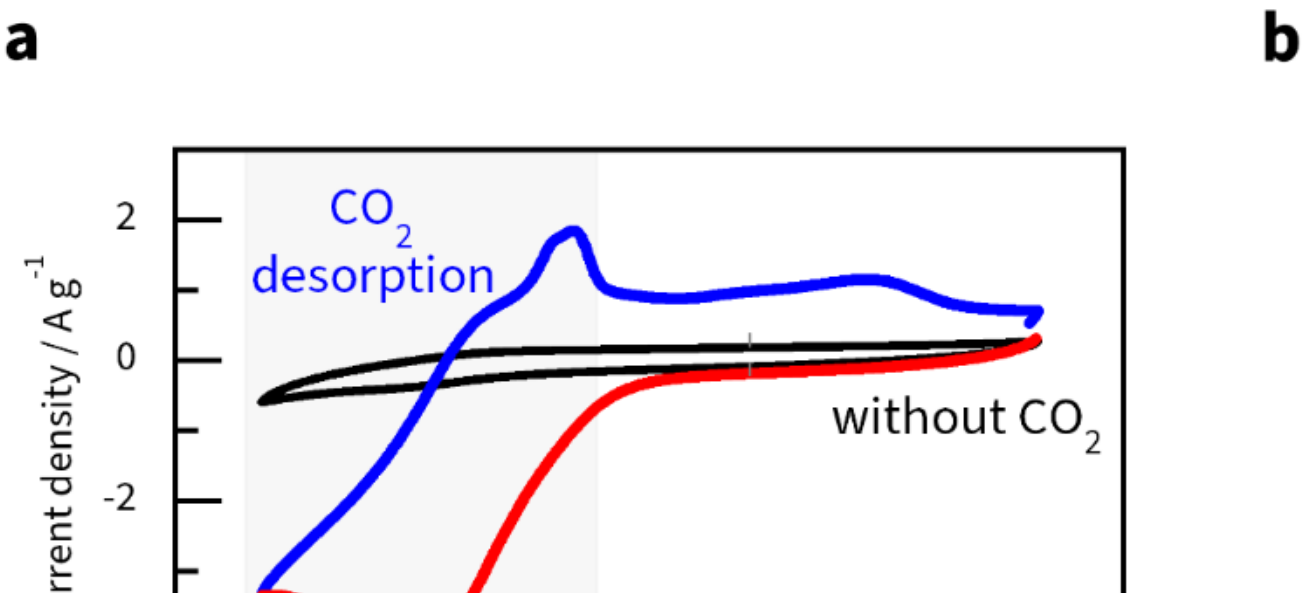

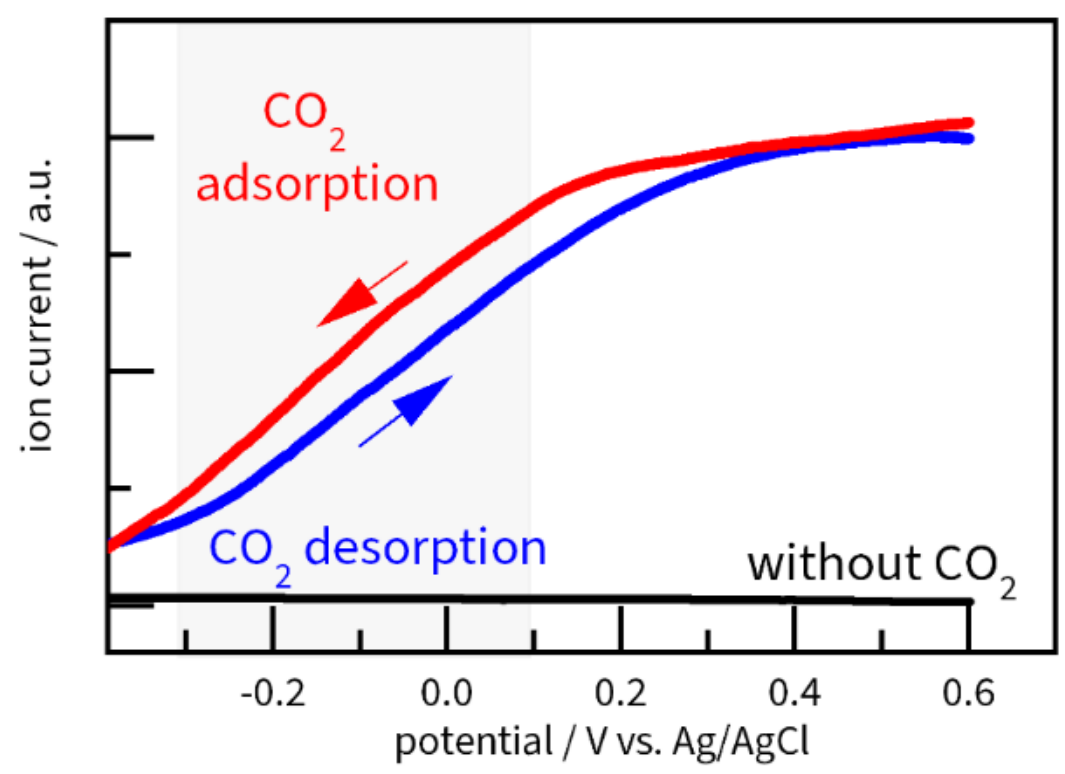

**Figure 3. (a)** CV curves recorded at $v$ = 10 mV s$^{-1}$ for the $Cu_3(HHTP)_2$-modified GC electrode: black line – in Ar-saturated electrolyte after stabilisation by cycling for two hours (without $CO_2$), red line – scan in the negative direction in $CO_2$-saturated electrolyte, blue line – scan in the positive direction in $CO_2$-saturated electrolyte. **(b)** DEMS ion current for $m/z$ = +44 $[CO_2]^+$ during CV at $v$ = 1 mV s$^{-1}$ for the $Cu_3(HHTP)_2$-modified GC electrode: red line – scan in the negative direction in $CO_2$-saturated electrolyte (showing adsorption from +0.2 V to −0.4 V); blue line – scan in the positive direction in $CO_2$-saturated electrolyte (showing $CO_2$ desorption from −0.3 V to +0.5 V); black line – in Ar-saturated electrolyte (without $CO_2$).

explore Co-, Ni-, and Cu-HHTP MOFs and found that among these, $Cu_3(HHTP)_2$ exhibits a capacity comparable to the long-known and optimised sorbents (Fig. 1, Table S1).

In this article, we introduce $Cu_3(HHTP)_2$ as the first redox-active MOF that electrosorbs $CO_2$ from an aqueous solution at ambient temperature. Using cyclic voltammetry (CV), electrochemical impedance spectroscopy (EIS), galvanostatic charge-discharge (GCD) tests, and differential electrochemical mass spectrometry (DEMS), we demonstrate the suitability of $Cu_3(HHTP)_2$ for the reversible $CO_2$ electrosorption. Using density functional theory (DFT) modelling, we provide atomistic insights into the electrosorption mechanism, revealing how the combination of redox properties of Cu and the aromatic system enables $CO_2$ chemisorption. Most importantly, this study demonstrates the key parameters that, when optimised, could lead to scalable and energy-efficient solutions for tackling global warming driven by $CO_2$ emissions.

## Results and discussion

### Structural characterisation

Since its initial synthesis in the 2010s, various $Cu_3(HHTP)_2$ morphologies, such as rod-, block- and agglomerated flake-like particles, have been reported.[53,54] Among these, the latter shows the best electrochemical performance in terms of capacitance.[53] Therefore, we synthesised the aforementioned flake-like morphology of $Cu_3(HHTP)_2$ (Fig. 2a), which was confirmed by scanning electron microscopy (SEM) analysis (Fig. 2b), elemental analysis, and specific surface area measurements (Figs. S1 and S2, Table S2 and S3). Simulated and measured powder X-ray diffraction (PXRD) confirms the crystallinity of the material (Fig. 2c). The synthesised $Cu_3(HHTP)_2$ appears as a layered material with hexagonal pores and slipped-parallel layer stacking (Figs. 2d and 2e).[55,56] Each Cu atom coordinates with four O atoms in the plane. The slipped-parallel layered stacking allows two-thirds of Cu atoms to coordinate with two additional O atoms from adjacent layers (as shown in Figs. 2e and 2f).[53,56,57]

Although model structures of MOFs are available,[58] specifically modelling the porous systems under electrochemical conditions is technically challenging due to the methodological limitations and computational costs that take the pore geometry into account. Combining density functional theory (DFT) calculations with molecular dynamics allows the study of physisorption processes,[59] alas, at a high cost. Thus, only a limited number of systems can be modelled. Incorporating $CO_2$ into such hybrid simulations complicates calculations even further. Hence, for this study, we have developed two-dimensional models of the $Cu_3(HHTP)_2$ pores (Figs. S3 and 2f) and run DFT calculations of electrochemical processes under constant potential and charge.[60] Figure 2f shows that in this model, $CO_2$ chemisorbs exclusively at the O sites of the MOF upon application of the potential.

Figures 2e and 2f illustrate that, as a result of slipped-parallel layer stacking and small O–O distance, only a fraction of O atoms are accessible for $CO_2$ chemisorption. In this case, all Cu atoms are completely sterically enclosed within the pore. Although some Cu and O atoms at terminal surfaces of $Cu_3(HHTP)_2$ microcrystals should be open for adsorption, this out-pore surface area is negligible compared to the in-pore surface area, even for the flake-like morphology.

### Electrochemical characterisation

Cyclic voltammetry measurements demonstrated the capacitive behaviour of $Cu_3(HHTP)_2$ after stabilisation by cycling for 2 hours in an Ar-saturated electrolyte (black curve in Fig. 3a and Fig. S4). Such behaviour is similar to the capacitive charging of thin layers and fine agglomerates of $Cu_3(HHTP)_2$,[61,62] discussed in Fig. S4.

In a $CO_2$-saturated solution, CV measurements showed a reversible redox process (red and blue lines in Fig. 3a) corresponding to $CO_2$ adsorption and desorption. According to the CV curves in Fig. 3a, the electrosorption on $Cu_3(HHTP)_2$ conceptually differs from electrosorption on anthraquinones.[63–65] For the latter, peaks in CV indicate reduction-oxidation of specific groups (with and without $CO_2$), which upon reduction can adsorb $CO_2$. In the absence of $CO_2$, $Cu_3(HHTP)_2$ shows no peaks, meaning that upon charging, electrons are distributed within the material (see the black curve in Fig. 3a), yet at a potential below +0.2 V, an electron transfer couples with $CO_2$ adsorption (see red curve in Fig. 3a).

To establish proof-of-concept electrosorption performance in aqueous media, long-term cycling stability, energy efficiency, and direct detection of $CO_2$ capture/release under simulated direct air capture conditions were conducted. Energy consumption and resistance characteristics were measured to further evaluate the electrochemical performance of $Cu_3(HHTP)_2$ in $CO_2$ capture/release. Galvanostatic charge-discharge (GCD) measurements were performed to quantify the energy requirements of the adsorption-desorption process (Fig. S5). The mean charge energy, determined by integrating the charge-discharge curves, was $8.5\cdot10^{-9}$ W·h, while the discharge energy was $7.2 \times 10^{-9}$ W·h, corresponding to an overall energy efficiency of 85%. In addition, interpretation of the electrochemical impedance spectroscopy (EIS) data showed that the process is limited by mass transport, i.e., diffusion of ions through pores, rather than electron transfer (Fig. S6 and Tab. S4).

To distinguish between $CO_2$ adsorption and pH-induced carbonate-related effects, we measured CV curves in $NaHCO_3$ and $Na_2CO_3$ electrolytes in the absence of $CO_2$ flow. This resulted in a noticeable decrease in current density (Fig. S7), suggesting that carbonate and bicarbonate ions do not induce the characteristic redox behaviour observed in $CO_2$-saturated conditions. In contrast, when gaseous $CO_2$ was introduced, the current density increased, and the redox peaks appeared, identical to those observed in Fig. 3a. These results confirm that the electrochemical response is caused primarily by the adsorption of gaseous $CO_2$ rather than carbonate species formed through $CO_2$ hydrolysis. Therefore, while pH controls carbon speciation in general DAC contexts, in the $Cu_3(HHTP)_2$ material's intrinsic molecular selectivity ensures that the sorption process is insensitive to bicarbonate or carbonate presence.

To qualitatively confirm $CO_2$ adsorption-desorption, we applied differential electrochemical mass spectrometry (DEMS) – a technique for monitoring gas adsorption and desorption processes by detecting molecular ion signals in real-time.[66] In this study, DEMS was used to monitor the $[CO_2]^+$ ion current at $m/z$ = +44[67,68] during a CV sweep from −0.4 V to +0.6 V (Fig. 3b and Fig. S8). It should be stressed that DEMS shows cumulative changes in the $CO_2$ sorption by the MOF, which, instead of reproducing discrete peaks, shows either a plateau (no sorption) or a slope (active adsorption or desorption). Moreover, the sweep rate is lower (1 mV $s^{-1}$) than one in the CV measurements (10 mV $s^{-1}$) due to the time resolution of DEMS.[69] Most importantly, control experiments without $CO_2$ flow showed no $CO_2$ signals, confirming that DEMS specifically detects adsorbed and desorbed $CO_2$ (Fig. 3b). During the cathodic scan (red line), the $CO_2$ signal remained relatively stable between +0.6 V and +0.2 V, indicating that $Cu_3(HHTP)_2$ does not adsorb $CO_2$ within this potential range. A rapid, linear decrease in detected $CO_2$ occurred from +0.1 V to −0.3 V, which implies the adsorption of $CO_2$ by the MOF. Active desorption occurred between −0.3 V and +0.5 V (blue line), in accordance with the CV in Fig. 3a. The agreement between CV and DEMS results suggests reversible $CO_2$ electrosorption on $Cu_3(HHTP)_2$ to happen most actively between –0.3 and +0.3 V vs. Ag/AgCl.

The $CO_2$ adsorption in a charged MOF is a slow process. Thus, to reach the maximum $CO_2$ capacity of Cu3(HHTP), we swept the potential from +0.6 V to −0.3 V and held it for a variable amount of polarisation time. Ten minutes were enough to saturate the MOF with $CO_2$ (Fig. 4a), i.e., to reach a constant $CO_2$ capacity value. The CV curve shows one cathodic (C1) and multiple (A1–

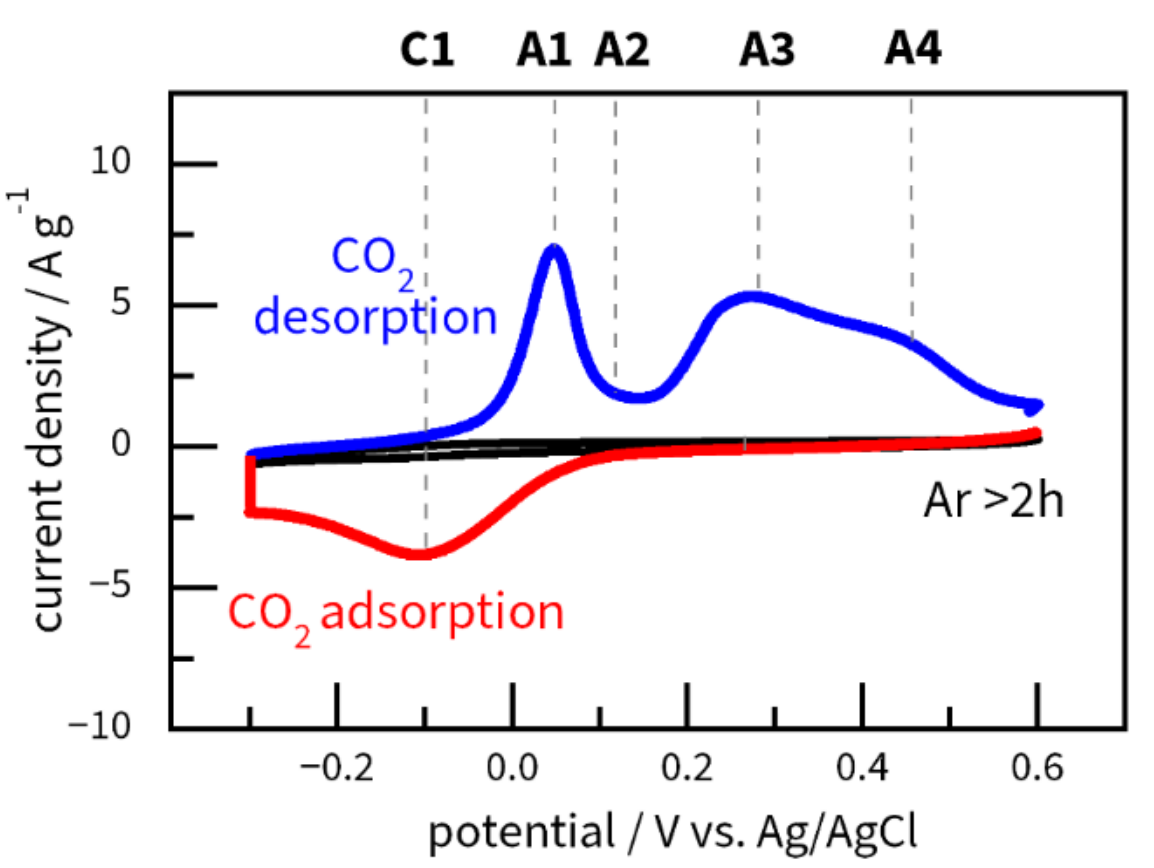

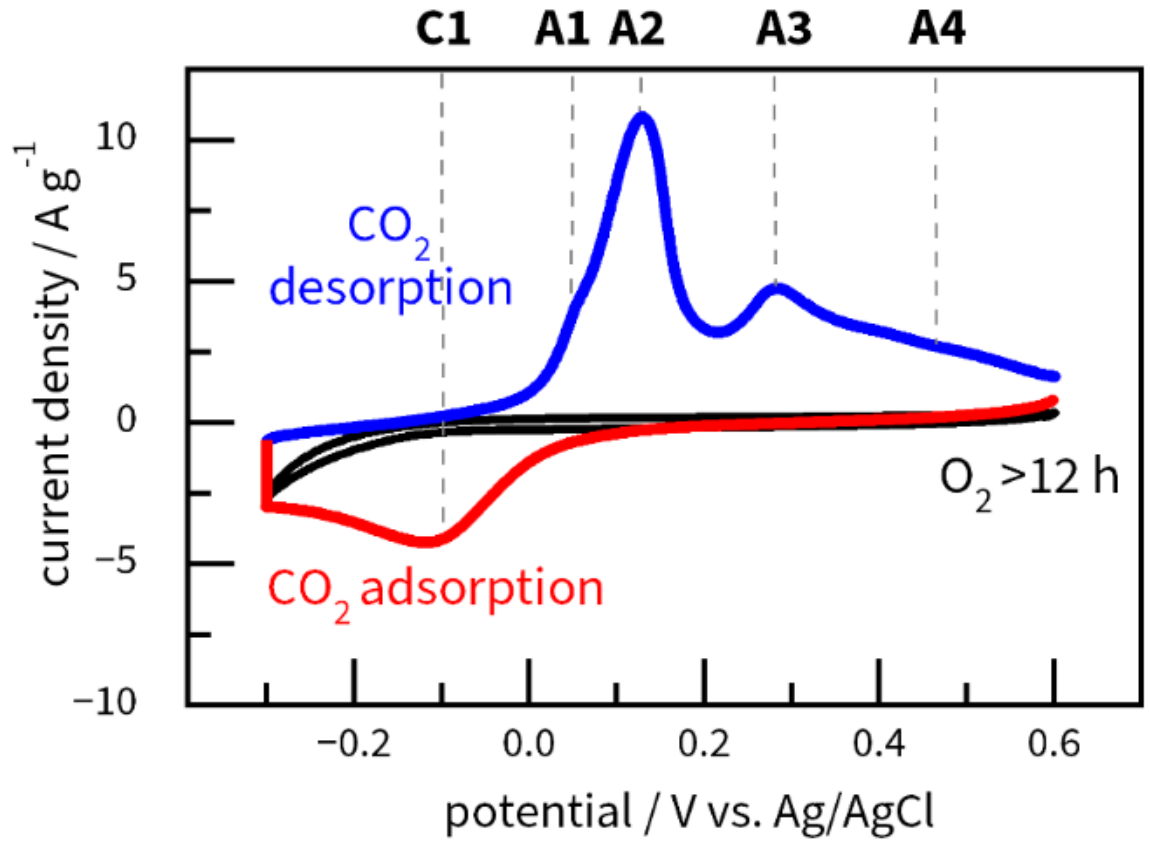

**Figure 4.** CV curves recorded at $v$ = 10 mV $s^{-1}$ for the $Cu_3(HHTP)_2$-modified GC electrode: **(a)** black line – in Ar-saturated electrolyte after stabilisation by cycling over two hours, red line – scan in the negative direction in $CO_2$-saturated electrolyte, blue line – scan in the positive direction in $CO_2$-saturated electrolyte, polarised for 10 min at −0.3 V. The C1-A1 peak separation is 150 mV. The $CO_2$ capacity of 2 mmol $g^{-1}$ was obtained for the material by considering the area under the blue line in the anodic region (2.3 V·A·$g^{-1}$) and subsequently subtracting the area under CV in Ar (0.007 A·V·$g^{-1}$); $C = (A_{CO2} - A_{Ar})\cdot(v\cdot F)^{-1}$ = (2.3 − 0.007)·(0.01·96485)$^{-1}$ = 2.38 mmol $g^{-1}$ (where $C$ is the $CO_2$ capacity of the MOF, $A_{CO_2}$ is the area under the positive-going scan measured in $CO_2$, $A_{Ar}$ is the area under the positive-going scan measured in Ar, $v$ is the scan rate, and $F$ is the Faraday constant). $CO_2$ capture capacity was estimated under the assumption that one electron induces adsorption-desorption of one $CO_2$ molecule.; **(b)** black line – scan in $O_2$-saturated electrolyte after 5000 cycles (over 12 hours), red line – subsequent scan in $CO_2$-saturated electrolyte in the negative direction, blue line – subsequent scan in $CO_2$-saturated electrolyte in the positive direction, polarised for 10 min at −0.3 V. The C1–A2 peak separation is 240 mV.

A4) anodic peaks with a separation of 150 mV between C1 and A1 peaks. If the process were purely thermodynamic, this separation would be directly related to the free energy of adsorption and desorption of around 14 kJ/mol. The amplitude of the peaks depends on the experimental conditions, like electrolyte composition, polarisation time, and sweep rate – very similar to anthraquinones.[39] From repeated experiments, we identified four anodic peaks at +0.05 V (A1), +0.13 V (A2), +0.27 V (A3) and +0.42 V (A4). Notably, the area under these peaks remains roughly the same in measurements run after keeping the MOF in Ar-saturated (Fig. 4a), $O_2$-saturated (Fig. 4b), and air-saturated (Fig. S9) electrolytes, which indicates the stability of $Cu_3(HHTP)_2$ in conditions close to ambient. Assuming that the current is linked to the $CO_2$ desorption as one electron per one $CO_2$ molecule, we report the capacity of 2 mmol of $CO_2$ per gram of $Cu_3(HHTP)_2$. One electron is needed to reduce $Cu^{2+}$, whereas adsorption happens on O sites surrounding the reduced $Cu^+$. This suggests a possibility for adsorption of more than one $CO_2$ molecule per electron.

Accordingly, we attribute the multiple anodic peaks (A1–A4) to the coupled oxidation–desorption process. As suggested by previous studies,[39,40] the neighbouring adsorbed $CO_2$ molecules repel each other. Removing the first $CO_2$ molecule, surrounded by adsorbed neighbouring $CO_2$ molecules, is energetically more favourable (corresponding to peak A1) as it eliminates repulsion. Further removal of the remaining $CO_2$ molecules requires more energy for desorption (corresponding to peaks A2–A4). In other words, we hypothesise that multiple oxidation peaks appear in this MOF because of numerous and distinct adsorption states, where $CO_2$ binding is affected by its surroundings. Notably, this study's peak separation (particularly between A1 and C1) is considerably smaller than those reported for other electrosorbents,[24,40,70] likely due to weaker $CO_2$ chemisorption.

To further validate the reversible $CO_2$ electrosorption performance beyond the standard three-electrode cell, we used an extended electrochemical flow battery setup with an electrode area of 9 $cm^2$ (Fig. S10a–c). In this system, $Cu_3(HHTP)_2$ was utilised to capture $CO_2$ directly from an airflow. While monitoring the system with a spectroscopic sensor, a distinct decrease in the $CO_2$ concentration was observed at negative potentials, indicating $CO_2$ capture (Fig. S10d). Upon switching to positive potentials, the $CO_2$ concentration increased, confirming the release process.

Altogether, CV, GCD, and DEMS measurements demonstrate that the redox behaviour of $Cu_3(HHTP)_2$ in $CO_2$-saturated electrolyte is directly associated with $CO_2$ electrosorption. This finding highlights the potential of similar redox-active MOFs in the research, understanding, optimisation, and application of $CO_2$ electrosorption for carbon capture.

**$CO_2$ electrosorption: Distinction from anthraquinones**

A previously known electrosorption mechanism involves the electrochemical reduction of an adsorption site, which alters its Lewis basicity, enabling $CO_2$ chemisorption.[71] Electrochemical reduction is usually denoted as step 'E', while chemisorption is denoted as step 'C'.[39] These two steps can occur in a concerted and sequential manner. For example, electrosorption on anthraquinones is thought to follow an 'ECEC' two-electron process, where two ketone moieties are reduced to alkoxide groups that bind two $CO_2$ molecules.[72] Reversing this process requires additional energy to break the formed bonds, resulting in a positive shift and separation of the anodic peaks. The data from DFT calculations reported in the literature supports this mechanism by linking binding energies to potential shifts.[40,41]

The MOF's resonance structure and +2 oxidation number of Cu imply that half of the O atoms in $HHTP^{3-}$ are alkoxides, while the other half are ketones. Thus, if a fully symmetric arrangement is assumed, the ratio between Cu atoms and ketones in the $CuO_4$ moiety equals 2. At first glance, the reduction of ketones could resemble the behaviour in anthraquinone, where $CO_2$ electrosorption follows a two-electron 'ECEC' process. Literature X-ray photoelectron spectroscopic data on $Cu_3(HHTP)_2$ suggests that $Cu^{2+}$ and $HHTP^{3-}$ are partially reduced. Our CV data for $Cu_3(HHTP)_2$ in Ar- and $CO_2$-saturated electrolytes shows a clear distinction from the features of (poly)anthraquinone.[39] The reduction curves in Figs. 3 and 4 show a singular reduction peak (C1), which the following mechanism can explain: when a $CO_2$ molecule binds to the MOF, it withdraws the electron density from the one-electron-reduced $Cu_3(HHTP)_2$. In a way, $CO_2$ reoxidises the MOF, which allows for its repeated reduction at C1, giving way to the adsorption of another $CO_2$ molecule until all four O sites in the $CuO_4$ moiety get occupied. Accordingly, as discussed above, the CV curve's oxidation side shows several peaks, indicating a sequential desorption.

**$CO_2$ electrosorption: New mechanism**

Using DFT calculations, we further elucidated the mechanism of $CO_2$ adsorption on $Cu_3(HHTP)_2$, which should be regarded as a qualitative guide along with experimental charge-discharge measurements (Fig. S5) providing the quantitative assessment of energy consumption. DFT results revealed that $CO_2$ binds exclusively to the oxygen sites when the MOF is reduced by one electron per Cu atom (Fig. S3). In this reduced state, the charge is evenly distributed between the $CuO_4$ moiety and the aromatic HHTP ligand (Fig. 5b). Copper gains approximately 0.1 electrons, maintaining its +2-oxidation state. In contrast, the HHTP ligands gain most of the added charge from the electron. This one-electron transfer resembles a capacitive charging as in graphitic materials with a fraction of the classical $Cu^{2+}/Cu^{+}$ reduction. Such DFT-based insight aligns with the observed capacitive charging from Fig. 3a and Fig. S4.

Upon adsorption, the $CO_2$ molecule withdraws 0.5 electrons from the nearest Cu atom, effectively integrating itself into the MOF structure without significantly altering the charge distribution in the HHTP ligands. The potential energy of adsorption of gaseous $CO_2$ onto the hydrated surface is −6 kJ $mol^{-1}$ (Fig. 5c). The lower absolute adsorption energy implies a lower cost for reversing the reaction. Recalculation into enthalpy gives an absolute value of −20 kJ $mol^{-1}$, which is notably lower when compared to other chemisorption mechanisms (Fig. 1 and Table S1).

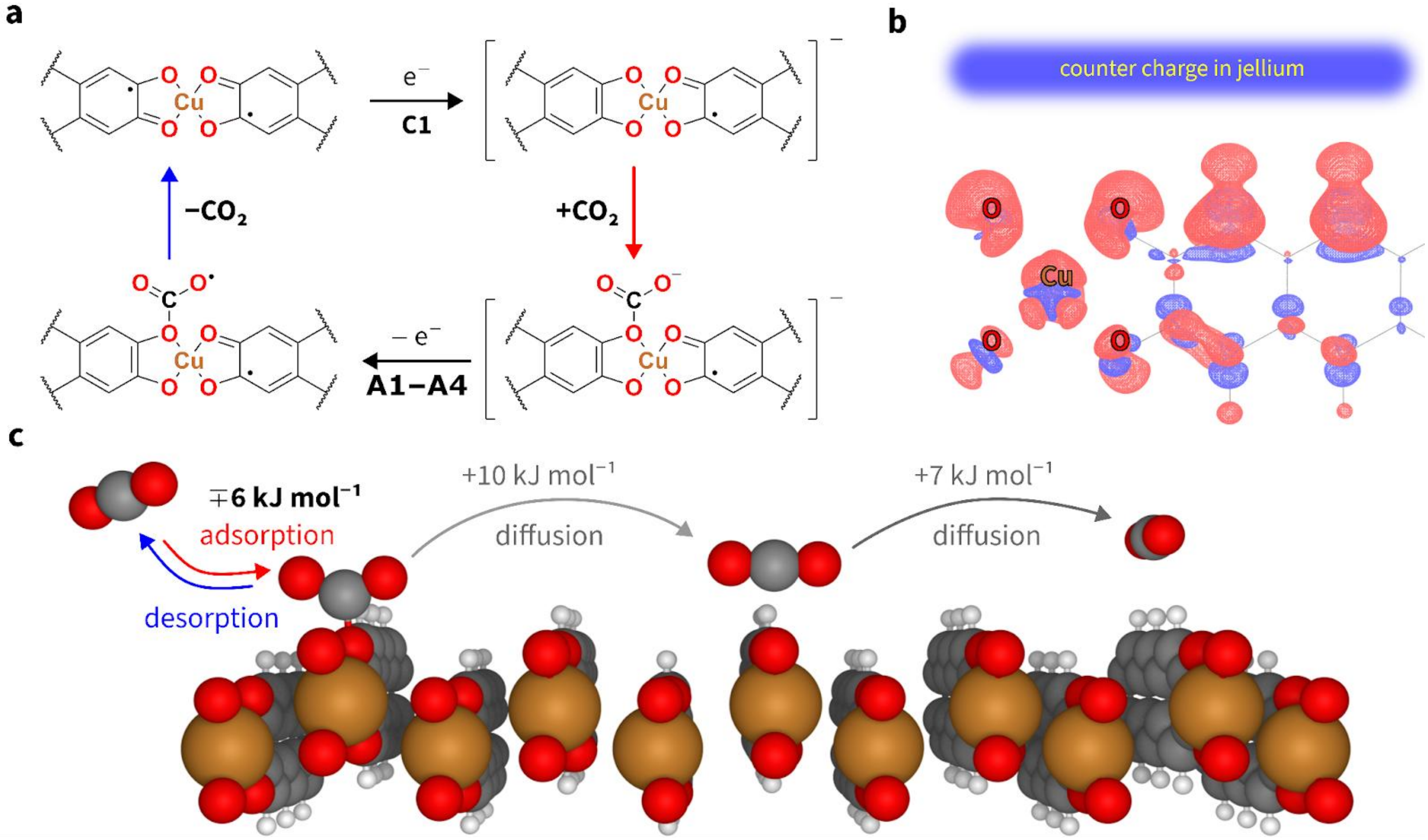

**Figure 5. (a)** Possible 'EC'-type mechanism of $CO_2$ electrosorption in $Cu_3(HHTP)_2$. A localised model of the MOF shows just one of many resonance structures with localised single- and double-bonds; **b)** isosurfaces of +0.002*e* (red) and −0.002*e* (blue) for charge density difference after adding 1*e* per Cu atom to a neutral 2D model of the MOF (see Figure S2). It shows an almost homogeneous distribution of the extra electron over the $CuO_4$ moiety and aromatic ring. The counter-charge was located in the 0.3 nm wide jellium region 0.6 nm above Cu atom; **c)** two-dimensional model of charged $Cu_3(HHTP)_2$ with $CO_2$ chemisorbed on the O site and physisorbed over Cu and aromatic rings. Numerical values represent potential adsorption and diffusion energies obtained from DFT calculations. The DFT analysis is based on a simplified 2D MOF model and is intended as a qualitative guide; quantitative energy consumption is provided by experimental charge–discharge data (Fig. S5).

DFT calculations and CV measurements show that $CO_2$ electrosorption on MOFs ($Cu_3(HHTP)_2$ in particular) differs from previously studied electrosorbents, such as anthraquinones.[39] First and foremost, pure $Cu_3(HHTP)_2$ does not show reduction-oxidation peaks. Secondly, DFT calculations reveal partial charge transfer to $CO_2$ during adsorption. Accounting for the latter, we can reevaluate the value of $CO_2$ capacitance from 2 to 4 mmol $g^{-1}$, around ¼ of the theoretical maximum of 14.5 mmol $g^{-1}$. As mentioned above, due to the steric hindrance caused by the stacked layers of the MOF structure, not all oxygen sites may be accessible (Figs. 2 and 5). Likewise, a $CO_2$ adsorption event on one of the O sites in a $CuO_4$ moiety could hinder the adsorption of subsequent $CO_2$ molecules on the neighbouring O site within the same pore. That might limit the practical $CO_2$ capacity to around ¼ of the theoretical maximum of 14.5 mmol $g^{-1}$. Such estimation agrees with the highest experimentally obtained $CO_2$ capacity of 2–4 mmol $g^{-1}$ (Fig. 4a). The sterical explanation also concurs with the work of Le et al., where sterically hindered phenazine macrocycles exhibit twice lower capacity (per electron) than molecular phenazine.[73]

### $CO_2$ electrosorption: Future work

The DFT results highlight the essential role of copper in facilitating reversible $CO_2$ capture. Copper is a donor of electron density that the $CO_2$ molecule can withdraw upon adsorption. Thus, these results point to future strategies to further enhance the capacity and efficiency of MOFs under realistic conditions. Such methods include: 1) selecting metals with redox properties that support reversible reduction, as copper does with its $Cu^{2+}/Cu^{+}$ states (Fig. 3); 2) choosing adsorption sites with suitable Lewis basicity, such as chalcogenides and pnictogens; and 3) optimising the aromatic counterparts to enhance both stability and porosity, thereby increasing capacity and improving mass transport. Beyond this fundamental understanding, we also addressed more practical questions presented below.

### $Cu_3(HHTP)_2$ stability

We assessed the stability of $Cu_3(HHTP)_2$ for $CO_2$ electrosorption through long-term CV experiments. In the range of −0.3 V to +0.7 V, the material demonstrates stability over 5000 cycles (Fig. S11). Copper underpotential deposition (Cu-UPD) tests (Fig. S12) confirm the absence of copper ions in the electrolyte after 1000 cycles. The actual degradation of $Cu_3(HHTP)_2$ happens below −0.4 V (Fig. S12). These findings suggest that $Cu_3(HHTP)_2$ remains stable and retains its $CO_2$ electrosorption capability over 24 h of $CO_2$ capture–release cycling.

### $M_3(HHTP)_2$ MOFs with different metals

Among the studied MOFs, only $Cu_3(HHTP)_2$ could electrosorb $CO_2$ within the potential range stable for aqueous solutions. When exposed to $CO_2$, neither $Ni_3(HHTP)_2$ nor $Co_3(HHTP)_2$ showed any signs of $CO_2$ adsorption in CV measurements (Fig. S15). These observations suggest that the copper centres in $Cu_3(HHTP)_2$ are involved in the electrosorption mechanism. Accordingly, we hypothesise that the metal's reduction potential should be in the optimal potential range for $CO_2$ electrosorption to serve as an electron donor to the adsorbing $CO_2$ molecule. That makes the $CuO_4$ moiety in $Cu_3(HHTP)_2$ unique and emphasises the need to explore similar MOFs with various tunable geometric and electronic features. For example, CoPc–$CuO_4$ and NiPc–$CuO_4$ also show the ability to electrosorb $CO_2$,[74] and suggest that among the conductive MOFs, there could be those with adsorption energy above –15 kJ mol$^{-1}$ and capacity over 5 mmol g$^{-1}$, i. e. in the desired region in Fig. 1a.

## Methods

### Synthesis of $Cu_3(HHTP)_2$

$Cu_3(HHTP)_2$ was synthesised using a literature procedure.[53] A solution of $Cu(NO_3)_2 \times 3H_2O$ (0.064 g, 0.265 mmol, 1.7 eq) and 28% aqueous ammonia (1.1 mL, 7.71 mmol, 50 eq) was prepared by dissolving them in Milli-Q water (1 mL). A dispersion of 2,3,6,7,10,11-hexahydroxytriphenylene (HHTP) was made separately by adding HHTP (0.051 g, 0.156 mmol, 1 eq) to Milli-Q water (4.1 mL). The copper solution was added dropwise to the HHTP dispersion in a 20 mL vial, and the reaction mixture was heated at 80 °C for 24 hours. The resulting product was collected by centrifugation, washed with water, ethanol, and acetone, and dried at 80 °C for 24 hours, giving black particles of $Cu_3(HHTP)_2$ (0.049 g, 75%). The resulting $Cu_3(HHTP)_2$ material was characterised using scanning electron microscopy (SEM), energy dispersive X-ray spectroscopy (EDX), and powder X-ray diffraction (PXRD).

### Physical characterisation

The $Cu_3(HHTP)_2$ MOF powder was secured onto stainless-steel scanning electron microscopy (SEM) stubs using adhesive high-purity carbon tabs. SEM images of the sample were obtained using a Tescan MIRA3 FEG-SEM, a high-performance field emission scanning electron microscope coupled with an Oxford Instruments X-maxN 80 energy dispersive X-ray spectroscopy (EDS) system for EDS acquisition and analysis. Imaging was conducted with a beam voltage of 5 kV with an In-Beam SE detector.

Powder X-ray diffraction (PXRD) data was collected over a 2θ range of 3–50° with a 0.050° step size under ambient conditions using Bruker D8 Advance diffractometer with LynxEye EX position sensitive detector. A homogenous sample was packed and flattened on a steel sample holder, 8.5 mm in height, with a sample reception Ø 25 mm. Computational structures used to produce the simulated PXRD patterns are available in Ref. [75]

$N_2$ adsorption isotherms were recorded at 77 K using an Anton Paar Autosorb iQ-XR instrument. Gas sorption analysis was performed for six samples. Prior to adsorption measurements, the samples were activated by heating under vacuum at 110 °C for 16 hours to remove residual atmospheric $CO_2$ and $H_2O$ from surfaces.

### Electrochemical characterisation

A glassy carbon (GC) working electrode (Origalys) with an area of 0.196 cm$^2$ was modified with a layer of $Cu_3(HHTP)_2$ suspension. The MOF powder was dispersed in a 0.5% solution of Nafion in isopropanol, prepared by adding 5% Nafion (10 µL) to isopropanol (90 µL). $Cu_3(HHTP)_2$ (1 mg) was added to 1 mL of the 0.5% Nafion solution in isopropanol, and the suspension was sonicated (NE00922, 40 kHz, 120 W) for 15 seconds. The suspension (5 µL) was dropcasted to the surface of the electrode and subsequently dried in the ambient air. The second layer was applied right after drying. The final $Cu_3(HHTP)_2$ loading on the electrode was 0.01 mg or 0.05 mg cm$^{-2}$.

The five-inlet electrochemical glass cell was used for CV, EIS and GCD experiments. CV measurements were performed in a three-electrode setup consisting of a GC working disk electrode (or Pt disk electrode, diameter 5 mm, Origalys), an Ag/AgCl (saturated KCl) reference electrode (or Pt wire electrode for EIS), and a GC rod counter electrode separated by a glass frit membrane in a five-inlet electrochemical cell. The electrolyte was 0.1 M sodium perchlorate ($NaClO_4$) solution prepared in Milli-Q water. CV scans were recorded at a sweep rate of 10 mV s$^{-1}$, typically in the potential ranges of −0.3 V to +0.6 V and −0.8 V to +0.8 V vs. Ag/AgCl, unless otherwise noted. The cell was saturated with pure Ar, $CO_2$, $O_2$ or air, and during the experiments, a continuous flow of corresponding gas was maintained above the solution to prevent contamination. All CV curves were recorded by a Multi Autolab M204 potentiostat controlled by the Nova v2.1 software of Metrohm. Experiments were reproduced at least three times.

### $CO_2$ gas flow cell measurements

Experiments were conducted in a flow electrochemical cell supplied with ambient air at the inlet. Two carbon paper (CP) electrodes with the area of 9 cm$^2$ each were employed; the working electrode was coated by drop-casting 10 mg of $Cu_3(HHTP)_2$ dispersed in 1 mL of 100% isopropanol, while the counter CP electrode was left unmodified. The electrodes were separated by a polyethersulfone membrane (Labbox MFPE-247-200). The outlet stream was monitored with an inline spectroscopic sensor Gas Lab GC018 to track $CO_2$ concentration. For adsorption, a potential of –0.3 V was applied for 300 s to saturate the working electrode with $CO_2$, followed by desorption at +0.5 V for 300 s

### Differential Electrochemical Mass Spectrometry measurements

During electrochemical measurements, DEMS was employed for *in-situ* monitoring of the adsorption and desorption of $CO_2$ on $Cu_3(HHTP)_2$. A Pfeiffer Prisma QMS 200 mass spectrometer equipped with a quadrupole detector and a secondary electron multiplier was coupled to a conventional three-electrode cell. A glassy carbon disk with a small cavity, approximately 1.5 mm in diameter, was used as the working electrode. A PTFE membrane (Gore-Tex) was positioned within this cavity, allowing the simultaneous acquisition of mass spectrometric and cyclic voltammograms with optimal sensitivity. The experiments were conducted at a scan rate of 1 mV $s^{-1}$ in a hanging meniscus configuration. The ion current for $m/z$ = +44 [$CO_2^+$] was tracked during potential sweeps between +0.6 V and −0.4 V. Experiments were reproduced at least three times.

#### Density Functional Theory Calculations

Density Functional Theory (DFT) calculations were performed using the Atomic Simulation Environment (ASE) version 3.23.0[76] and the GPAW package version 24.1.0[77] to model the $CO_2$ adsorption on $Cu_3(HHTP)_2$. The RPBE exchange-correlation functional,[78] with D4 dispersion correction,[79] was employed alongside a projector augmented-wave method to describe core and valence electrons. Spin-polarisation was turned on, and the magnetic moment of ±1 was preset on $Cu^{2+}$ ions, with alternating signs for distinct layers. The Brillouin zone was sampled with four k-points in the MOF plane and two k-points in perpendicular directions. $CO_2$ adsorption was modelled in finite-difference mode using a solvated jellium model,[60] with a constant counter charge of +1e per Cu atom positioned above the MOF surface. An implicit water layer was included to simulate solvation effects. Adsorption energies were calculated for three sites (above Cu atoms, O atoms, and aromatic rings) with optimisation of atomic positions until the residual forces were below 0.1 eV Å$^{-1}$. Enthalpy, entropy, and free energy values were evaluated at ideal gas and harmonic approximations through a vibrational analysis implemented in the ASE thermochemistry module.

## Conclusions

This study demonstrated that conductive and redox-active metal-organic frameworks (MOFs) can reversibly electrosorb $CO_2$ under mild conditions. $Cu_3(HHTP)_2$, in particular, operates efficiently in the aqueous solution at ambient temperature and in the presence of oxygen. That represents a significant advancement in $CO_2$ capture technology, as the studied class of materials – conductive and redox-active MOFs – can operate under the conditions required for direct air capture. However, to assess its practical application, a thorough technoeconomic and stability analysis is required. Our findings reveal that the mechanism of electrosorption in $Cu_3(HHTP)_2$ involves a synergistic interplay between copper centres, oxygen sites, and aromatic ligands. The redox activity of copper centres modulates the affinity at oxygen sites, enabling controlled adsorption and desorption cycles without the electrochemical reduction of $CO_2$. This mechanism is distinct from previous electrosorbents, such as polyanthraquinone, in which the process does not involve metal centres.

Based on the modelled mechanism, this study suggests optimising variable metal centres, adsorption sites, and aromatic structures to design new conductive redox-active MOFs capable of reversible $CO_2$ electrosorption. That represents a conceptual shift in $CO_2$ capture strategies, moving towards integrating conductive MOFs with specific redox and adsorption functionalities. Optimising the parameters identified in this study could lead to the development of the first generation of electrolytic cells for direct air capture of $CO_2$, ultimately reducing atmospheric $CO_2$ levels.

## Author contributions

V.I., V.G., and N.K. conceived the project, designed the experiments and supervised the work. I.V. carried out electrochemical measurements. I.V. collected and interpreted the electrochemical data. N.Z. and A.K. synthesised the materials. K.M., A.C.F., and J.W.G. provided preliminary samples, while V.I., R.C., and V.K. performed the DFT calculations. T.T., H.Q., and P.C. ran physicochemical analyses. V.G., E.H., R.R., and J.S.-G. provided DEMS analysis. A.F., E.H., R.R., K.M. and J.S.-G. refined the manuscript. V.I., I.V., N.K., V.G., and N.Z. co-wrote the manuscript.

## Conflicts of interest

There are no conflicts to declare.

## Data availability

The data supporting the findings of this study are available within the article and its Supplementary Information files. The computational simulation data is available as an ASE database from Zenodo at https://www.doi.org/10.5281/zenodo.14216403. Any other data supporting this study's findings are available from the corresponding author upon reasonable request.

## Acknowledgements

This work was supported by the Estonian Ministry of Education and Research (TK210), the Estonian Research Council (STP52 MOB3JD1208, PUTJD1245 and PUTJD1244), Ministerio de Ciencia, Innovación y Universidades (PID2022-137350NB-I00 and PID2022-138491OB-C32 (MCIN/AEI /10.13039/501100011033 / FEDER, UE)), and the European Cooperation in Science and Technology Innovation Grant (COST CIG 18234, NanoCatML). Computational results were obtained using the UT Rocket High-Performance Computing Center of the University of Tartu.

Supporting information

for

# Demonstrating Electrochemical $CO_2$ Capture with Redox-Active Metal-Organic Frameworks

*Iuliia Vetik,*[a] *Nikita Žoglo,*[b] *Akmal Kosimov,*[c] *Ritums Cepitis,*[c] *Veera Krasnenko,*[d]
*Huilin Qing,*[e] *Priyanshu Chandra,*[e] *Katherine Mirica,*[e]
*Ruben Rizo,*[c] *Enrique Herrero,*[c] *Jose Solla-Gullón,*[c]
*Teedhat Trisukhon,*[f] *Jamie W. Gittins,*[f] *Alexander C. Forse,*[f]
*Vitali Grozovski,*[a] *Nadezda Kongi,*[a*] *Vladislav Ivaništšev.*[a,g*]

[a] Institute of Chemistry, University of Tartu, Tartu 50411, Estonia
[b] RedoxNRG OÜ, Narva-Jõesuu 29021, Estonia
[c] Institute of Electrochemistry, University of Alicante, Apdo. 99, 03080, Alicante, Spain
[d] Institute of Physics, University of Tartu, Tartu 50411, Estonia
[e] Department of Chemistry, Burke Laboratory, Dartmouth College, New Hampshire 03755, United States
[f] Yusuf Hamied Department of Chemistry, University of Cambridge, Cambridge CB2 1EW, UK
[g] Department of Chemistry, University of Latvia, Riga LV-1004, Latvia

**Table S1.** Comparison of experimental capacity and adsorption enthalpy values for selected $CO_2$ sorbents. Theoretical capacity values per molar mass of the sorbent are given in parentheses.

| Sorbent material | Capacity [mmol $g^{-1}$] | Conditions | Adsorption enthalpy [kJ $mol^{-1}$] | Ref. |
|---|---|---|---|---|
| **Absorption** | | | | |
| Oxide: CaO | 17.9 | any | −172 | 1 |
| Monoethanolamine 30 wt% | 2.5 (8) | 100% $CO_2$, 1 atm | −82 | 2 |
| Ionic liquid: [hmim][$Tf_2N$] | 1.2 | 100% $CO_2$ 13 atm | −13 | 3 |
| **Adsorption** | | | | |
| MOF: SIFSIX-3-Cu | 1.2 | 400 ppm $CO_2$ | −54 | 4 |
| Zeolite: 13-XPEI | 1.2 | 100% $CO_2$, 1 atm | −35 | 5 |
| **Electrosorption** | | | | |
| Poly(1,4-anthraquinone) | 8.4* (9.7) | $CO_2$ sat. org. electrolyte | −86 | 6 |
| MOF: $Cu_3(HHTP)_2$ | 0.8–2.4* (14.5) | $CO_2$ sat. aq. electrolyte | −20** | This work |

* Estimated from CV data.

** Calculated at the DFT level *via* vibrational analysis.

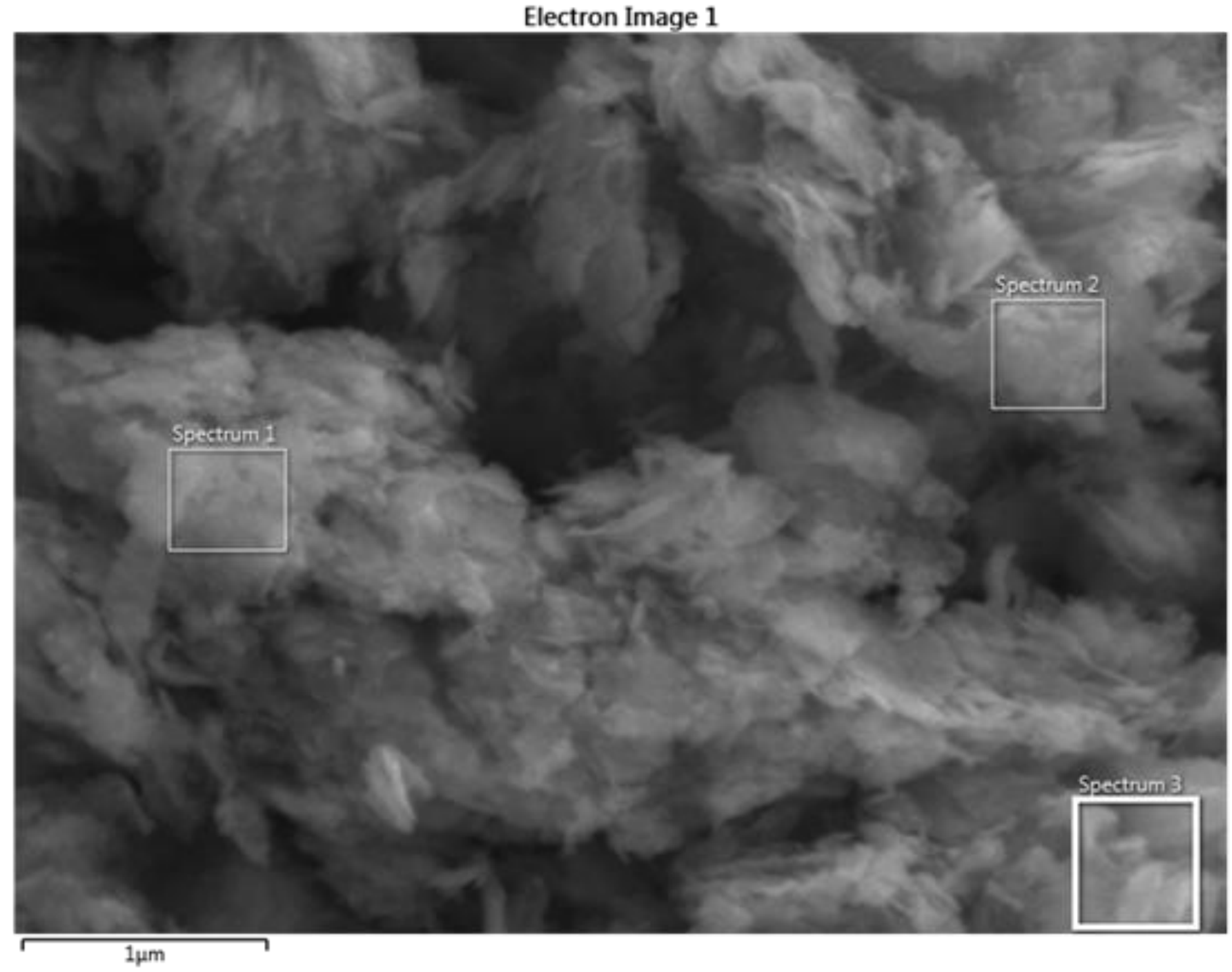

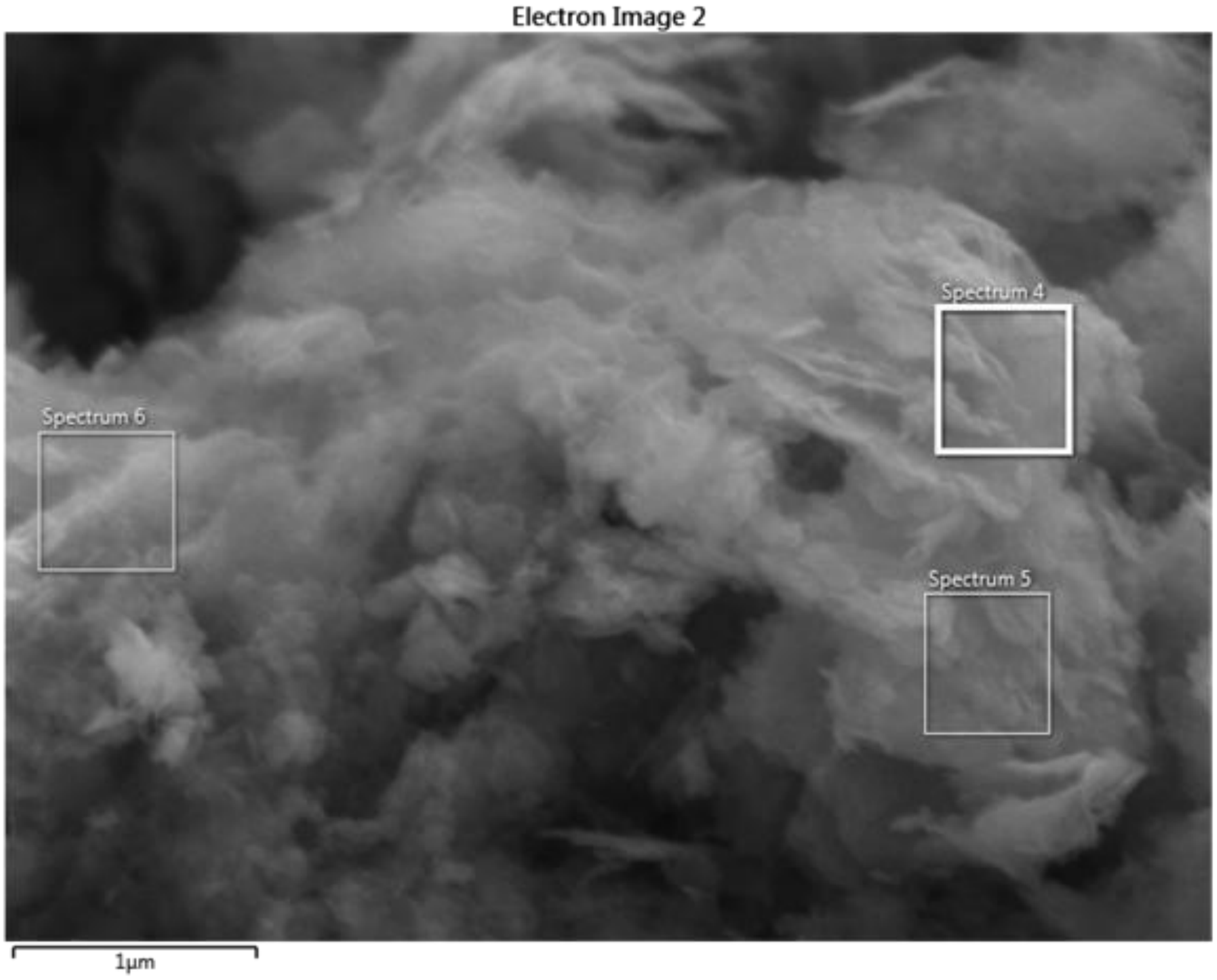

**Figure S1.** Scanning electron microscope (SEM) images obtained for $Cu_3(HHTP)_2$ during the energy dispersive X-ray (EDX) microanalysis with areas of analysis indicated by squares.

**Table S2.** The surface elemental composition of the $Cu_3(HHTP)_2$ sample obtained by the energy dispersive X-ray microanalysis. Some nitrogen content arises from the aqueous ammonia used in the synthesis and has been observed previously.[7]

| **Spectrum** | **C** | **O** | **N** | **Cu** |
|---|---|---|---|---|
| **1** | 53.4 | 16.1 | 2.4 | 28.1 |
| **2** | 53.7 | 15.1 | 2.1 | 29.2 |
| **3** | 52.1 | 16.6 | 2.7 | 28.6 |
| **4** | 55.5 | 14.8 | 1.6 | 28.2 |
| **5** | 54.4 | 14.2 | 1.2 | 30.2 |
| **6** | 53.5 | 18.5 | 2.4 | 25.6 |
| **Average** | **53.8** | **15.9** | **2.1** | **28.3** |
| **Error** | 1.1 | 1.6 | 0.6 | 1.5 |
| **Calculated[7]** | **52.3** | **23.1** | **0** | **23.1** |

**Table S3.** Elemental analysis for C, H, and N content of the $Cu_3(HHTP)_2$ sample and inductively coupled plasma optical emission spectroscopy (ICP-OES) results for Cu content. Some nitrogen content arises from the aqueous ammonia used in the synthesis and has been observed previously.[7]

| **Element** | **Calculated[7] (wt%)** | **Reference[7] (wt%)** | **Average found (wt%)** |
|---|---|---|---|
| **C** | 52.3 | 48.9 | **41.8** |
| **H** | 1.5 | 2.4 | **3.3** |
| **N** | 0 | 2.8 | **4.3** |
| **Cu** | 23.1 | 21.7 | **18.5** |

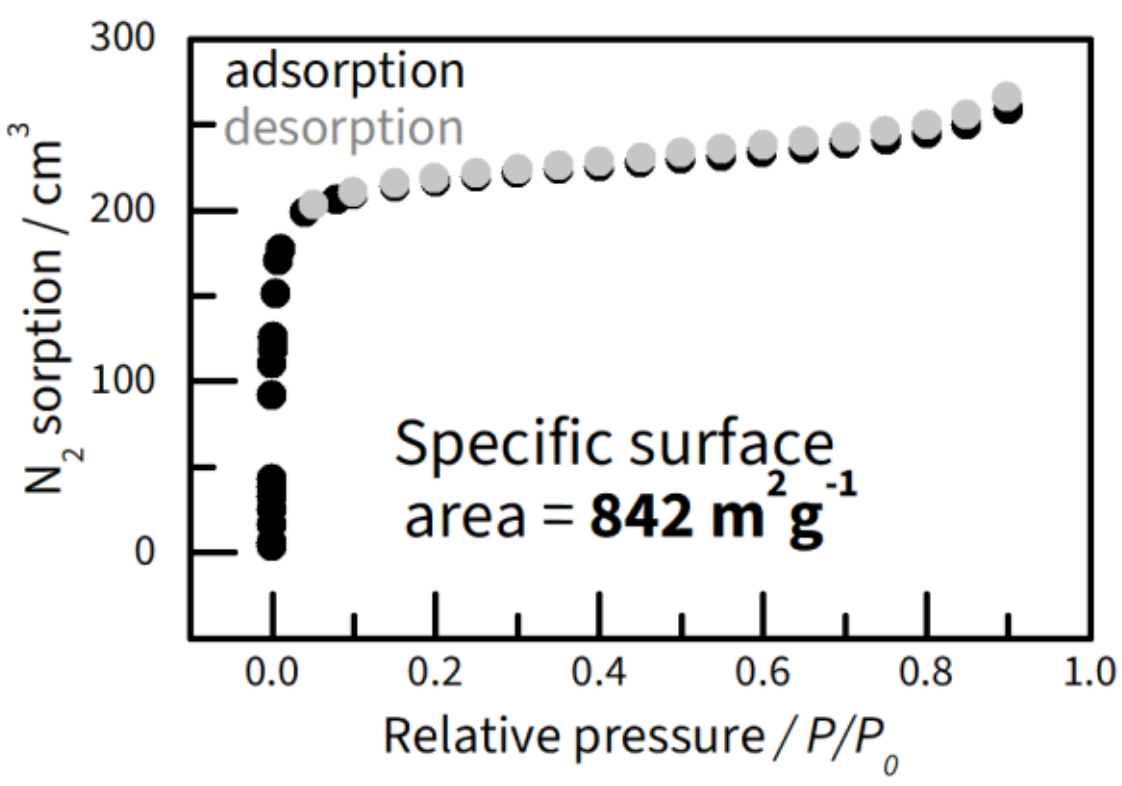

**Figure S2**. $N_2$ sorption isotherm for $Cu_3(HHTP)_2$.

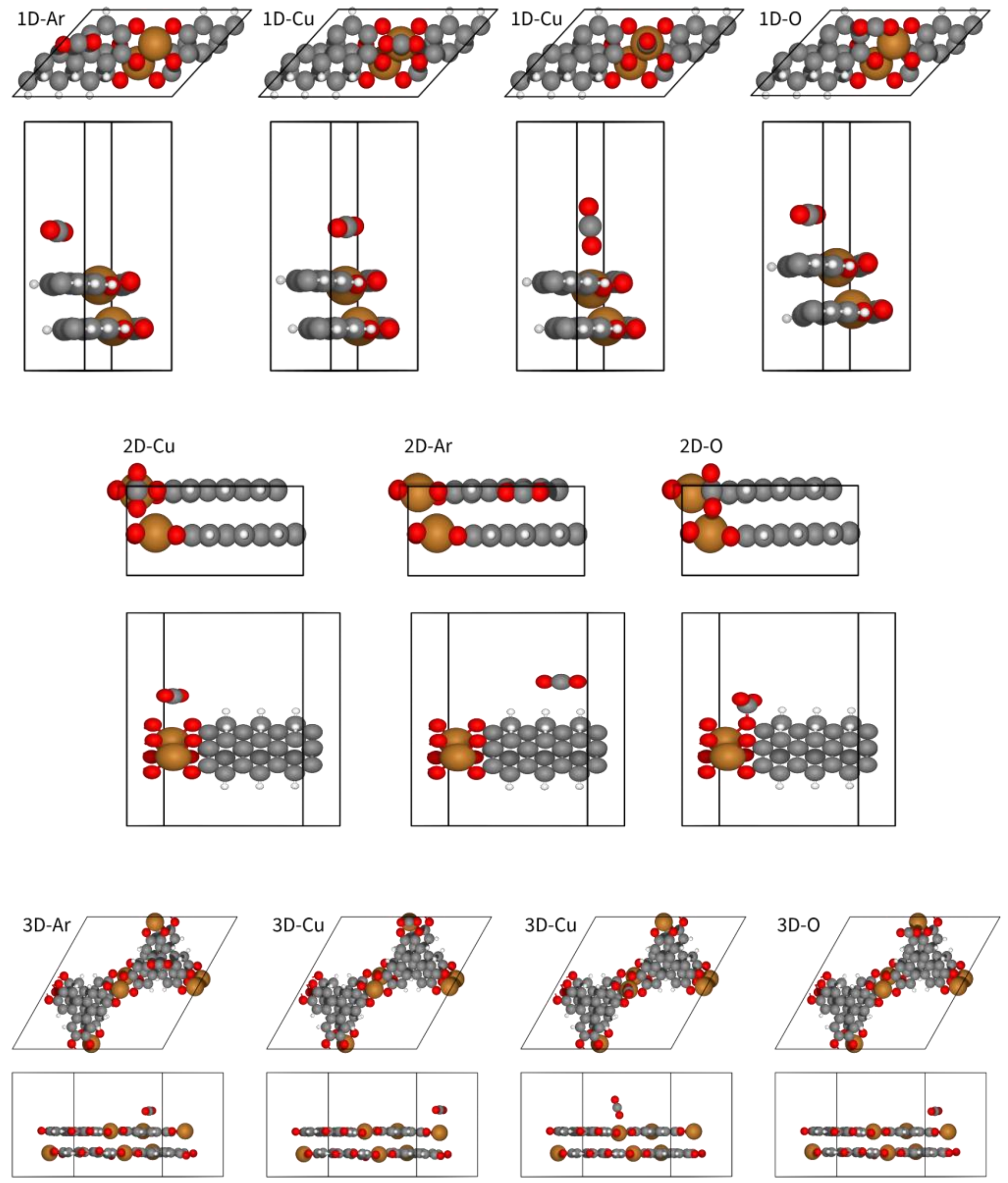

**Figure S3.** Top and side views on three models (**1D**, **2D**, and **3D**) of Cu-HHTP with the $CO_2$ molecule adsorbed at three sites (above **Cu**, **O**, **Ar**omatic system) and in two positions (adsorption through C or O in $CO_2$). The top view is labelled and situated above the side view of each model. The $CO_2$ molecule is seen in all side views, where the bent geometry of $CO_2$ indicates chemisorption (in the **2D-O** model). One- and two-dimensional (1D and 2D) models represent in-pore and terminal adsorption sites of a model MOF with features of Cu-HHTP: $CuO_4$ unit and three aromatic rings. The three-dimensional (3D) model represents terminal adsorption sites of $Cu_3(HHTP)_2$. In each model, there are two layers in the unit cell. For computational details, see the "Density Functional Theory (DFT) calculations" section below. All computational data are available in the format of the Atomic Simulation Environment database from Zenodo at DOI:10.5281/zenodo.14216403.

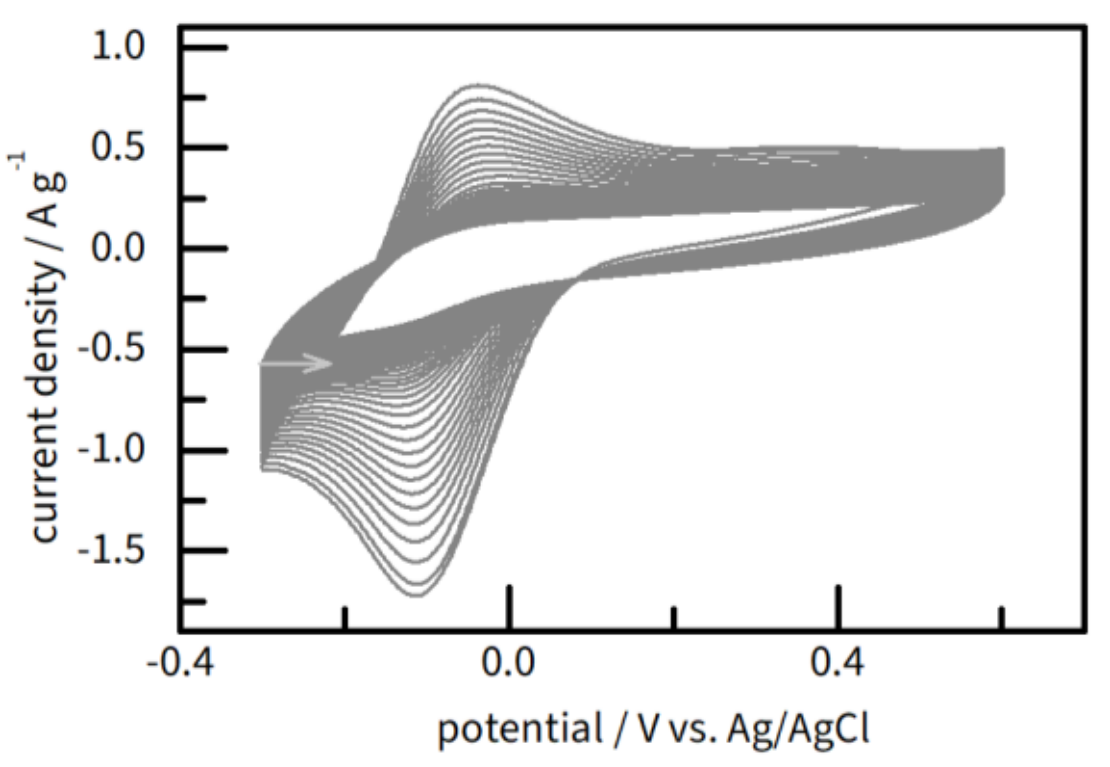

**Figure S4**. $Cu_3(HHTP)_2$ stabilisation in Ar-saturated 0.1 M $NaClO_4$ electrolyte. Cyclic voltammetry (CV) curves were recorded over 40 cycles for $Cu_3(HHTP)_2$-modified GC electrode at a scan rate of 10 mV $s^{-1}$.

The observed change in electrochemical behaviour is similar to previously reported dependences of redox activity on thickness,[8] and microstructure of $Cu_3(HHTP)_2$.[9] Zhao et al. showed that with a decrease of $Cu_3(HHTP)_2$ film thickness, the electrochemical behaviour changes from redox-active to capacitive, which is in line with decreasing the charge transfer resistance in impedance spectra.[8] Similarly, Gittins et al. showed that for finer microstructure grinding, the electrochemical behaviour was found to be limited by the ion mass transport resulting in lower current densities.[9] The mass-transport lamination is apparent for $Cu_3(HHTP)_2$ in this work (Fig. S6). Thus, it is possible to interpret the results in Fig. S4 as a structural transformation of $Cu_3(HHTP)_2$.

The disappearance of the cathodic peak at −0.16 V in Fig. S4 is unrelated to oxygen, as it diminishes with time, even in oxidative conditions in electrolytes saturated with air and $O_2$. The electroreduction of oxygen is expected to happen at lower potentials.[10] The observed change is also unrelated to the reduction of Cu, which potentially could remain from synthesis or form by MOF-s decomposition. Thus, the redox activity of pure MOF in the Ar-saturated 0.1 M $NaClO_4$ electrolyte (as well as in $NaHCO_3$, $Na_2SO_4$, and NaCl) is probably due to the restructuring, as suggested above.

Redox peaks observed in Fig. S4 were reported in previous works.[11–13] X-ray photoelectron spectroscopy revealed simultaneous reduction of $Cu^{2+}$ to $Cu^{+}$ and quinoid (C=O) to benzenoid (C−O) group.[12,13] XPS analysis of $Cu_3(HHTP)_2$ of variable morphology showed that predominant oxidation state of copper is +2,[14] and charge of −3 on HHTP,[15] due to 1:1 ratio between C−O to C=O. Notably, in 1 M KOH aqueous electrolyte, the discussed reduction is reversible by 99% after 1000 galvanostatic charge and discharge cycles.[16]

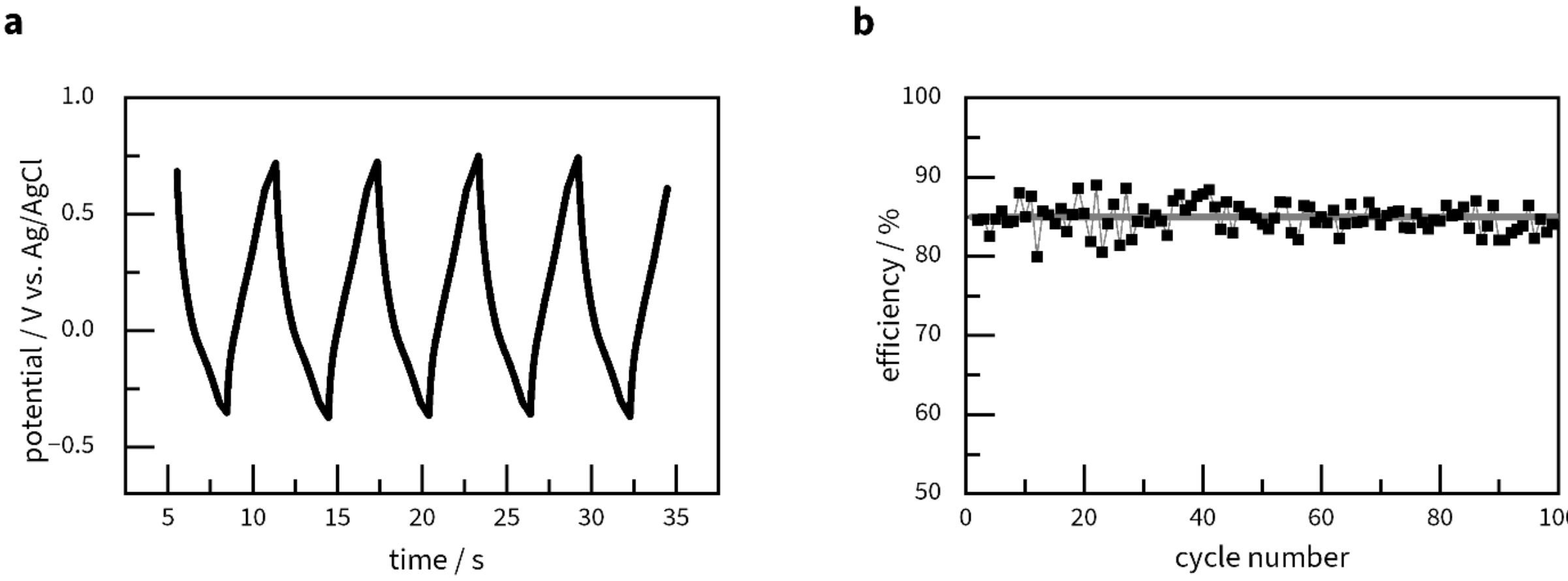

**Figure S5. (a)** First cycles of the galvanostatic charge/discharge (GCD) experiment for the $Cu_3(HHTP)_2$-modified GC electrode in a $CO_2$-saturated 0.1 M $NaClO_4$ electrolyte. The mean charge energy, determined by integrating the charge-discharge curves, was $8.5{\cdot}10^{-9}$ W·h, while the discharge energy was $7.2{\cdot}10^{-9}$ W·h. **(b)** The energy efficiency of $Cu_3(HHTP)_2$ calculated over 100 charge-discharge cycles, with the mean value of 85% represented by the grey line.

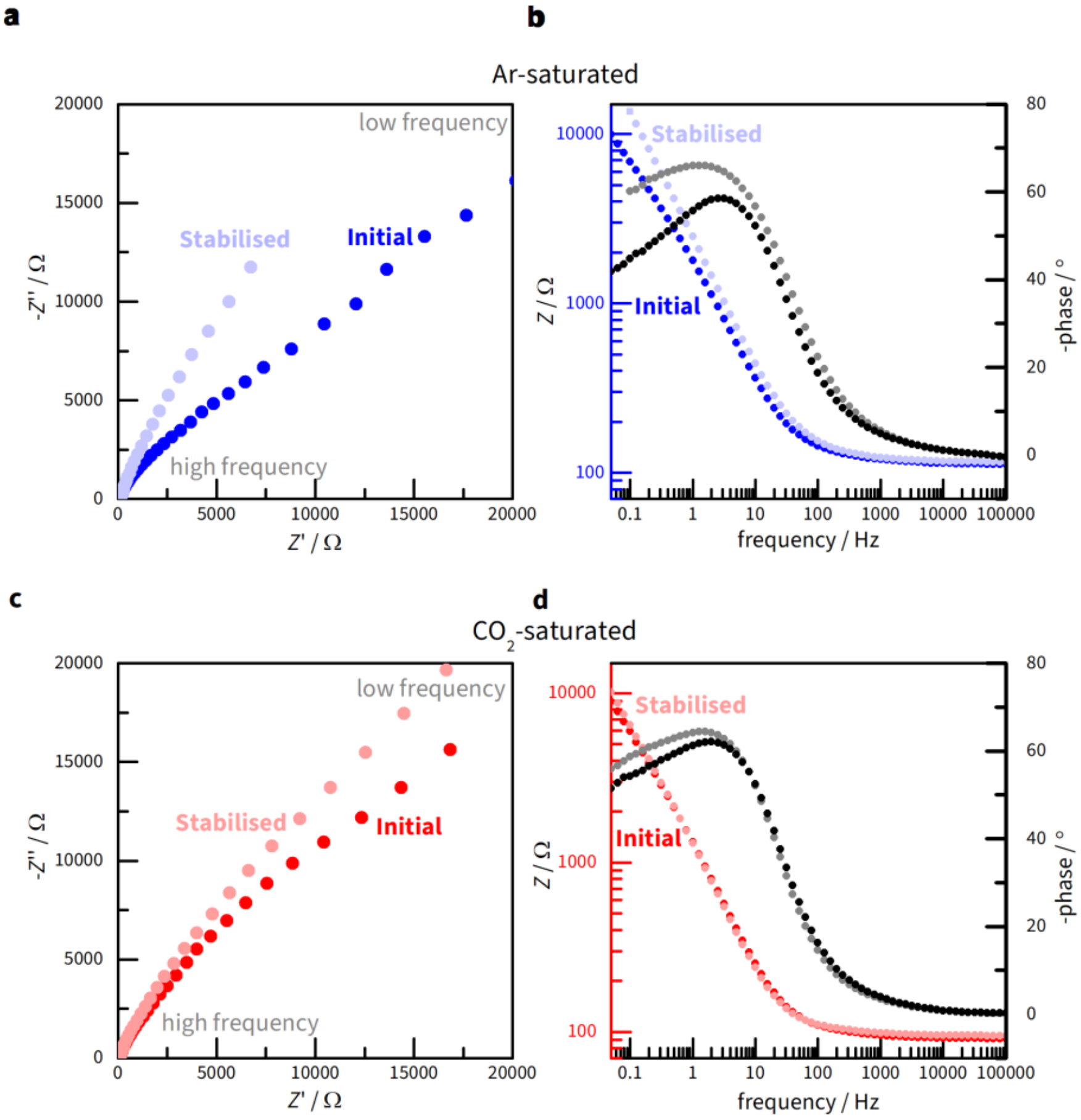

**Figure S6.** Electrochemical impedance spectroscopy (EIS) experiments performed with a $Cu_3(HHTP)_2$-modified GC electrode in Ar-saturated and $CO_2$-saturated 0.1 M $NaClO_4$ electrolyte. **(a) and (c)** Nyquist plots and **(b) and (d)** Bode plots. Darker points represent experimental data (blue: Ar-saturated electrolyte; red: $CO_2$-saturated electrolyte) for unstabilized $Cu_3(HHTP)_2$, while lighter points correspond to stabilized MOF.

At −0.3 V, the impedance data fit well to a model featuring a Warburg element, indicating that mass transport or diffusion-like processes play an important role in the system. The nearly negligible charge-transfer resistance ($R_{ct}$ = 6 μΩ) suggests that electron transfer through the $Cu_3(HHTP)_2$ interface is extremely facile under these conditions, making the primary rate limitation more associated with diffusion of ions or $CO_2$ within the porous framework. The double-layer capacitance ($C_{dl}$ = 32–68 μF) aligns with typical interfacial values for such conductive MOF-based electrodes, while the Warburg coefficient ($Y_0$ = 89–180 μMho $s^{0.5}$) captures the characteristic $\omega^{-0.5}$ dependence of diffusion-controlled impedance. Overall, the strong Warburg component suggests that at lower frequencies (down to 0.1 Hz) the system's response is dominated by the transport of adsorbate or ions in the MOF structure rather than by charge-transfer kinetics at the electrode surface.

**Table S4.** Electrochemical impedance spectroscopy (EIS) fitting results

| Potential / V vs. Ag/AgCl | Circuit | Electrolyte resistance / Ω | Capacitance / µF | $Y_0$ / µMho·$s^{0.5}$ | Resistance / µΩ | Capacitance / mF | $\chi^2$ |
|---|---|---|---|---|---|---|---|
| Initial in Ar-saturated electrolyte | | | | | | | |
| -0.145 | R = 119 Ω<br>C = 31.6 µF<br>W<br>Y0 = 167 µMho*s^(1/2) | 119 | 32 | 167 | - | - | 0.42 |
| Stabilised in Ar-saturated electrolyte | | | | | | | |
| -0.145 | R = 119 Ω<br>C = 10.5 µF<br>Y0 = 96.6 µMho*s^N | 119 | 11 | 97<br>(α = 0.7) | - | - | 0.09 |
| -0.145 | R = 119 Ω<br>C = 31.7 µF<br>W<br>Y0 = 88.8 µMho*s^(1/2) | 119 | 32 | 89 | - | - | 0.55 |
| -0.145 | R = 119 Ω<br>C = 25.2 µF<br>W<br>Y0 = 132 µMho*s^(1/2)<br>C = 214 µF | 119 | 25 | 132 | | 0.2 | 0.3 |
| -0.145 | R = 119 Ω<br>C = 31.7 µF<br>W<br>Y0 = 88.8 µMho*s^(1/2)<br>R = 6.00 µΩ | 119 | 32 | 89 | 6 | - | 0.55 |
| Initial in $CO_2$-saturated electrolyte | | | | | | | |
| -0.3 | R = 95.7 Ω<br>C = 57.5 µF<br>W<br>Y0 = 181 µMho*s^(1/2) | 95 | 58 | 181 | - | - | 0.22 |
| Stabilised in $CO_2$-saturated electrolyte | | | | | | | |
| -0.3 | R = 95.0 Ω<br>C = 53.8 µF<br>Y0 = 172 µMho*s^N | 95 | 54 | 172<br>(α = 0.57) | - | - | 0.15 |
| -0.3 | R = 95.0 Ω<br>C = 68.1 µF<br>W<br>Y0 = 154 µMho*s^(1/2) | 95 | 68 | 154 | - | - | 0.3 |
| -0.3 | R = 95.0 Ω<br>C = 65.0 µF<br>W<br>Y0 = 166 µMho*s^(1/2)<br>C = 3.62 mF | 95 | 65 | 166 | - | 3.6 | 0.25 |
| -0.3 | R = 95.0 Ω<br>C = 68.1 µF<br>W<br>Y0 = 154 µMho*s^(1/2)<br>R = 6.36 µΩ | 95 | 68 | 154 | 6 | - | 0.3 |

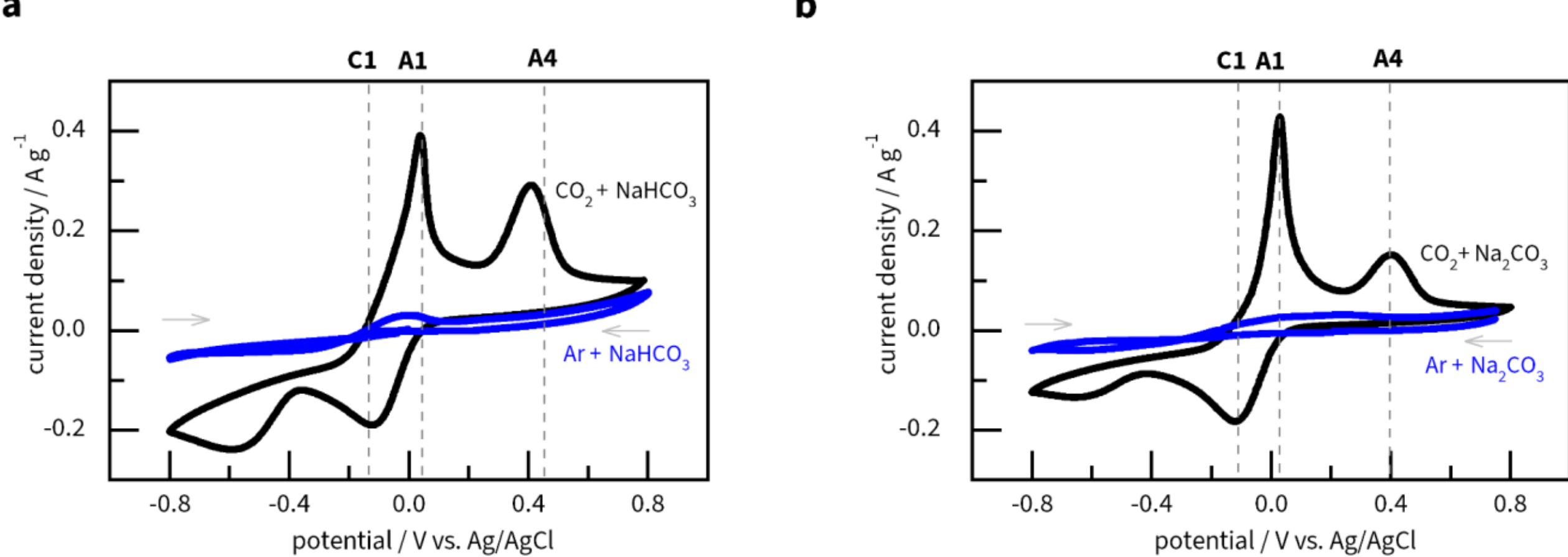

**Figure S7.** CV curves recorded for $Cu_3(HHTP)_2$-modified GC electrode at $\nu$ = 10 mV $s^{-1}$ in 0.1 M $NaClO_4$ electrolyte. **(a)** blue line – Ar-saturated electrolyte with 20 mM of $NaHCO_3$, black – $CO_2$-saturated electrolyte with 20 mM of $NaHCO_3$; **(b)** blue line – Ar-saturated electrolyte with 20 mM of $Na_2CO_3$, black – $CO_2$-saturated electrolyte with 20 mM of $Na_2CO_3$.

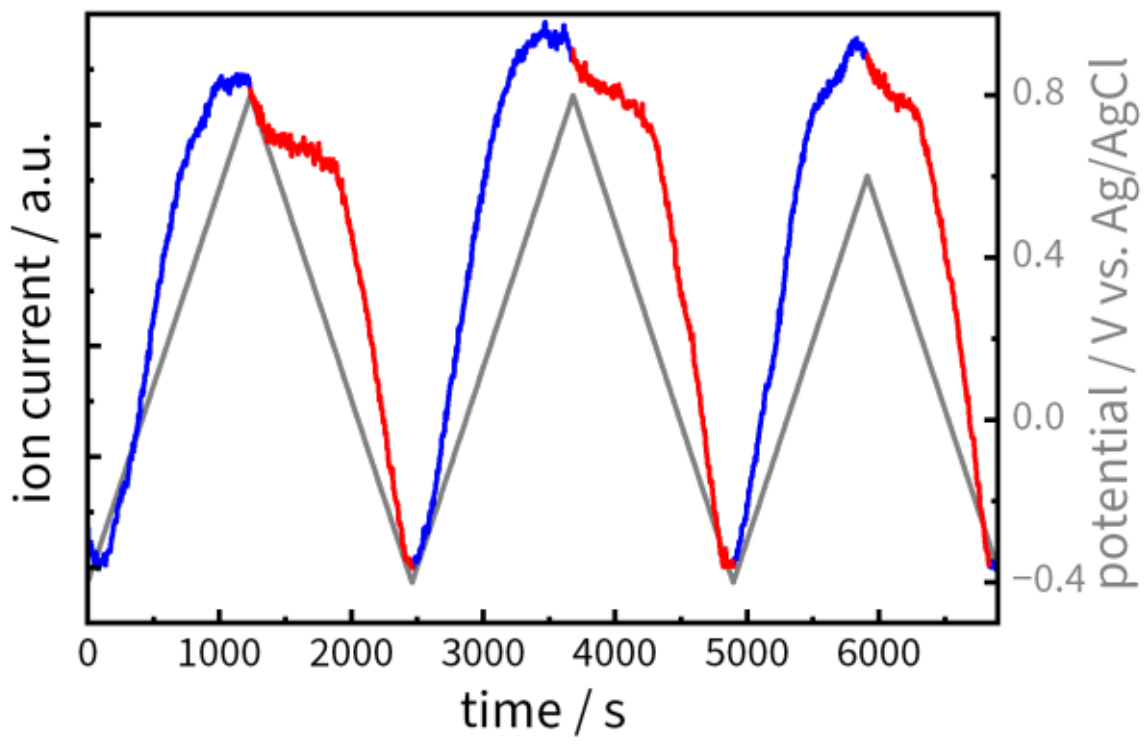

**Figure S8.** DEMS: ion current of *m/z* = +44 $[CO_2]^+$ during CV at ν = 1 mV $s^{-1}$ for the GC-$Cu_3(HHTP)_2$ working electrode in the presence of $CO_2$: red line – negative going scan, blue line – positive going scan, with the corresponding working electrode potential (right axis).

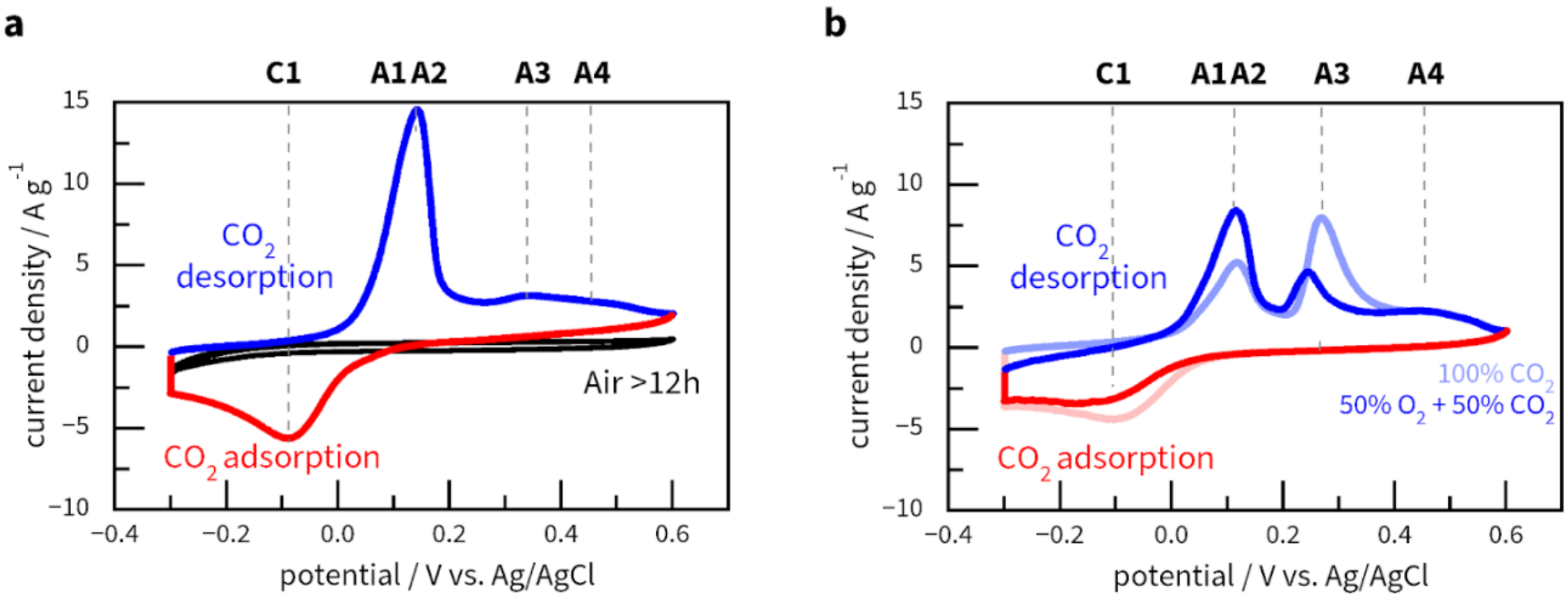

**Figure S9.** CV curves recorded at $v$ = 10 mV s$^{-1}$ for the $Cu_3(HHTP)_2$-modified GC electrode in 0.1 M $NaClO_4$ electrolyte: **(a)** The curves are measured after 5000 cycles in the air-saturated electrolyte. The black line is recorded in the air-saturated electrolyte, while the red and blue lines correspond to scans in electrolyte re-saturated with $CO_2$ and after polarization for 10 min at −0.3 V: the red line represents the negative direction, and the blue line represents the positive direction; **(b)** the pale red and pale blue lines correspond to scans in $CO_2$-saturated electrolyte after polarization for 10 min at −0.3 V. The red and blue lines correspond to scans in (50% $CO_2$ + 50% $O_2$)-saturated electrolyte after polarization for 10 min at −0.3 V: the red line represents the negative direction, and dark blue line represents the positive direction.

It is notable that 10 min polarization at −0.3 V is not enough for reaching saturation, i.e. the full sorption capacity, in the (50% $CO_2$ + 50% $O_2$)-saturated electrolyte, differently from the $CO_2$-saturated electrolyte. That is due to lower concentration of $CO_2$, which is also the reason why the shape of C1 peaks differ for these electrolytes. Insufficient saturation results in higher amplitude of A1/A2 peak for the (50% $CO_2$ + 50% $O_2$)-saturated electrolyte similarly to the higher amplitude in Fig. 3 in comparison to Fig. 4. Longer polarization and higher $CO_2$ concentration are needed for the appearance of A3 and A4 peaks. Still, the amount of desorbed $CO_2$ is comparable and, thus, it can be concluded that $O_2$ does not impede the $CO_2$ electrosorption.

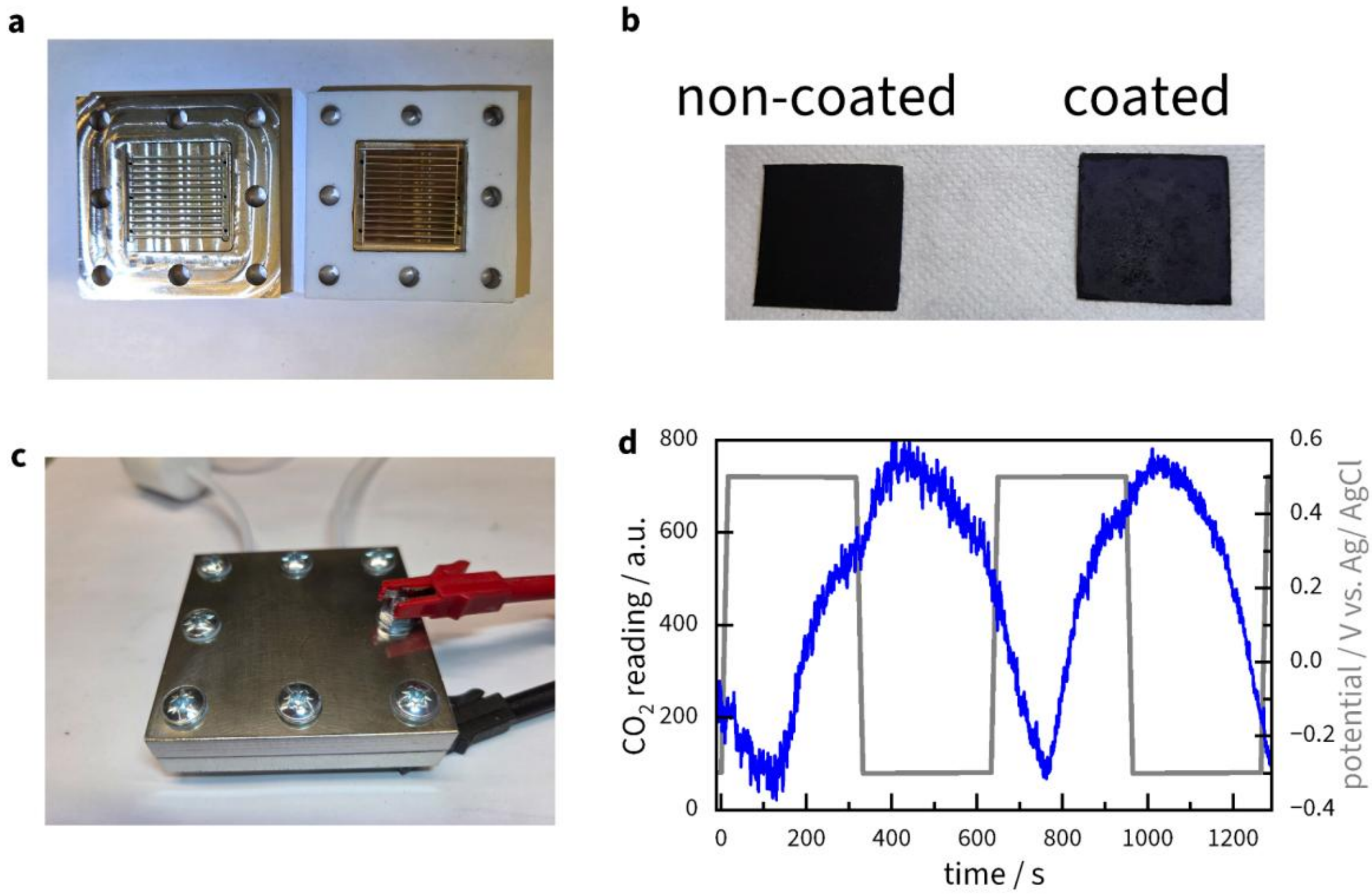

**Figure S10: (a)** Disassembled electrochemical gas flow cell; **(b)** carbon paper electrode, both uncoated (left) and coated with $Cu_3(HHTP)_2$ (right); **(c)** fully assembled electrochemical gas flow cell; **(d)** $CO_2$ concentration at the cell outlet (blue line), measured downstream of a carbon paper electrode loaded with 10 mg of $Cu_3(HHTP)_2$, where $CO_2$ was adsorbed/desorbed upon air introduction at the inlet. The corresponding working electrode potential is shown on the right axis. Duration of 300 seconds per half-cycle was found to be optimal based on the dynamic response of the system. At shorter cycle times, the $CO_2$ adsorption–desorption transitions would be incomplete, leading to attenuated amplitudes and underestimation of capture–release efficiency. Extending the cycle length to longer times would not further increase uptake/release, indicating that equilibrium was already reached well within 300 s.

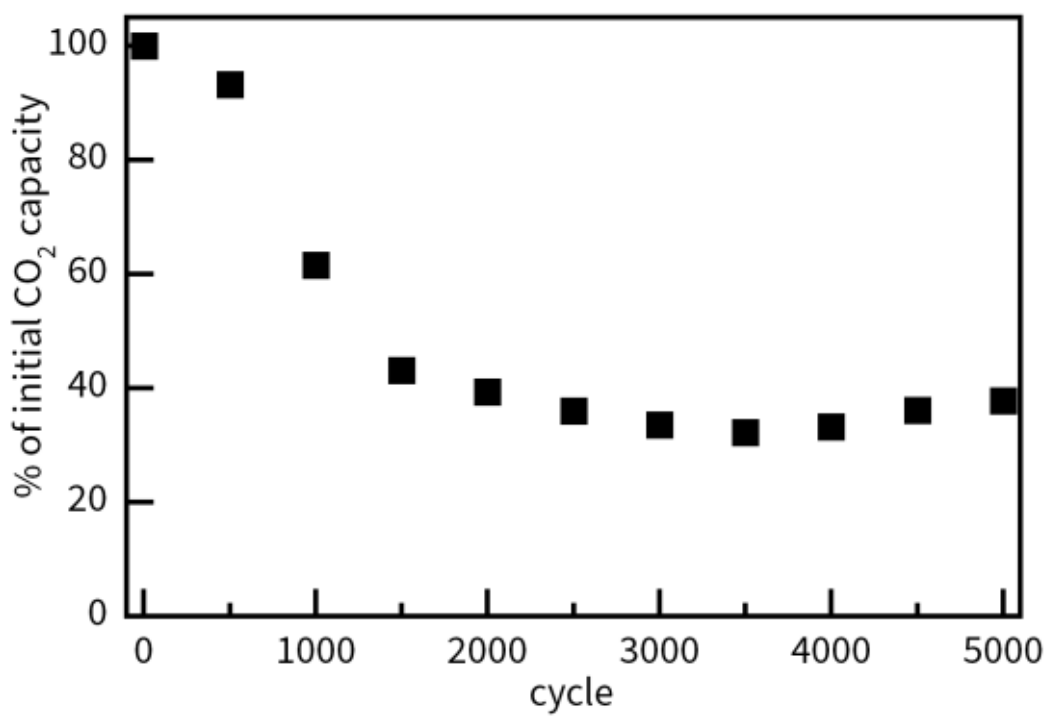

**Figure S11.** The drop in $CO_2$ capacity was studied during long-term cycling of the GC-$Cu_3(HHTP)_2$ electrode in a $CO_2$-saturated 0.1 M $NaClO_4$ electrolyte, within a potential range of −0.3 V to +0.7 V. The capacity reaches a plateau at approximately 30% after 2000 cycles.

After every 100 cycles at a scan rate of 200 mV $s^{-1}$, the electrode was polarized for 1 minute to ensure $CO_2$ saturation. CV was then recorded at a scan rate of 10 mV $s^{-1}$. The $CO_2$ capacity of the material was determined by integrating the area under the CV curve in the anodic region and subtracting the corresponding area under CV in an Ar atmosphere, using the equation:

$$C = (A_{CO_2} - A_{Ar}) \cdot (v \cdot F)^{-1}$$

where $C$ is the $CO_2$ capacity of the MOF, $A_{CO_2}$ is the area under the positive-going scan measured in $CO_2$, $A_{Ar}$ is the area under the positive-going scan measured in Ar, v is the scan rate, and $F$ is the Faraday constant. $CO_2$ capture was estimated under the assumption that 1 electron induces adsorption-desorption of one $CO_2$ molecule.

Herewith, the experimentally observed lowering of $CO_2$ capacity from 2 to 0.8 mmol $g^{-1}$ after 2 hours of cycling could be related to clogging of pores. These numbers indicate a potential for improvement through further optimisation of redox-active MOFs. Namely, adjusting the polarisation times and conditions needed to saturate and desaturate MOFs, i.e., respective $CO_2$ capture and release.

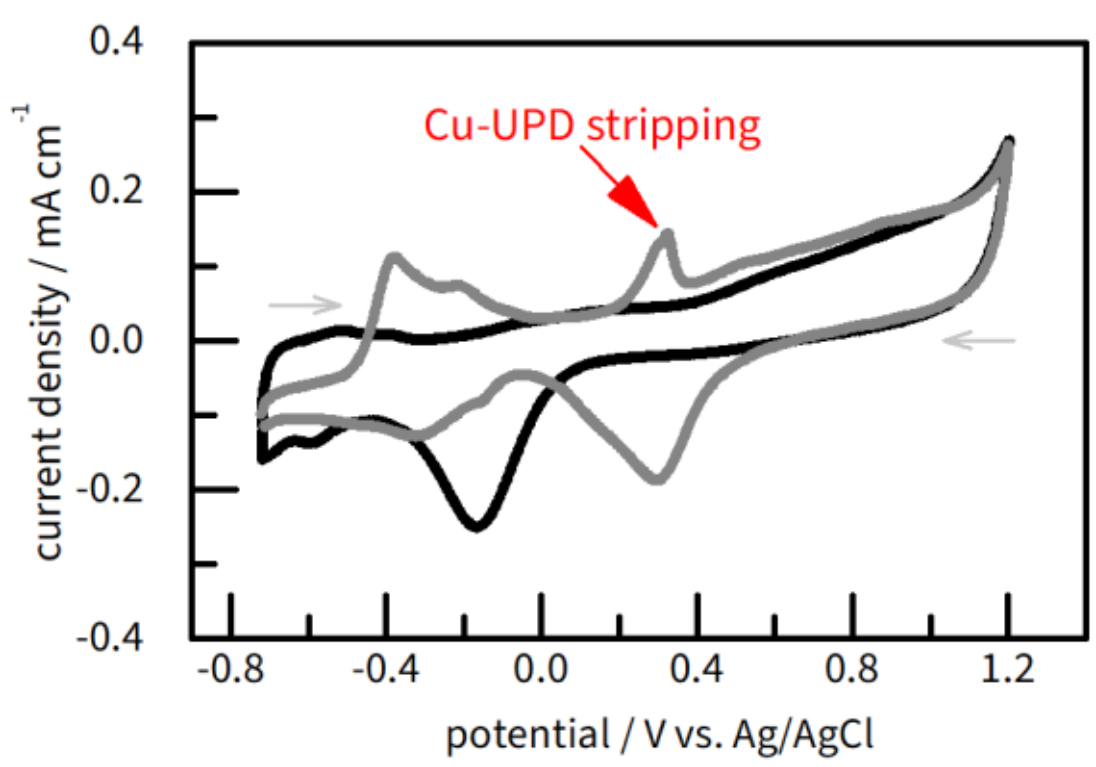

**Figure S12.** Cyclic voltammetry (CV) curves for copper stripping from Pt electrode, recorded at scan rate of 50 mV $s^{-1}$, after $Cu^{2+}$ underpotential deposition (UPD) from electrolyte extracted after experiments with $Cu_3(HHTP)_2$. The UPD was run for 20 cycles in the potential window from 0 to −0.72 V. This experiment tests contamination of the electrolyte by $Cu^{2+}$ due to decomposition of $Cu_3(HHTP)_2$ MOF. Grey line – CV curve in the electrolyte obtained after cycling $Cu_3(HHTP)_2$ for 1000 cycles in the potential range of −0.8 V to +0.8 V, anodic peak at 0.35 V on a grey line indicates Cu-UPD stripping.[17] Black line – CV in the electrolyte obtained after cycling $Cu_3(HHTP)_2$ for 5000 cycles in the potential range of −0.3 V to +0.7 V. The absence of the Cu-UPD peak on the black line proves that $Cu_3(HHTP)_2$ does not release $Cu^{2+}$ during operation within the window used to study the $CO_2$ electrosorption.

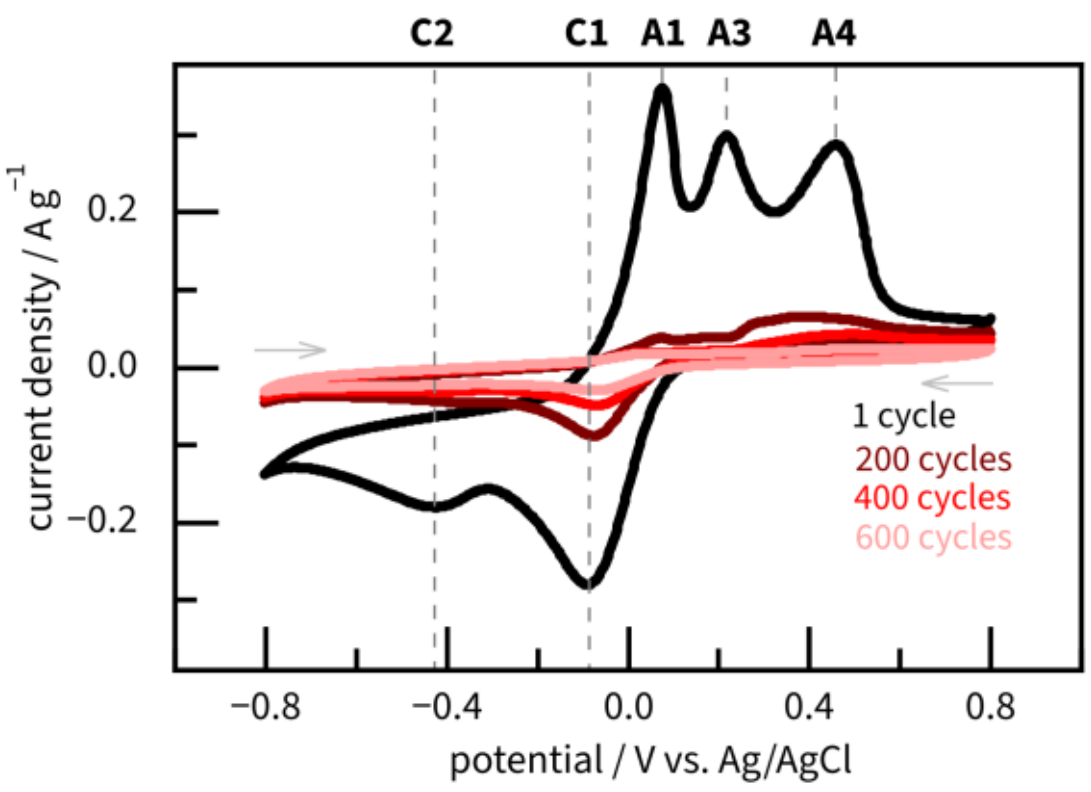

**Figure S13.** CV curves recorded after long-term cycling at $v$ = 10 mV s$^{-1}$ in $CO_2$-saturated 0.1 M $NaClO_4$ electrolyte for GC-$Cu_3(HHTP)_2$ in the potential range from −0.8 V to +0.8 V: black – after 1 cycle, dark brown – after 200 cycles, red – after 400 cycles, light pink – after 600 cycles. The expected product of decomposition is $Cu_2O$.[18] Besides, the decomposition releases $Cu^{2+}$ ions to the electrolyte (see Fig. S12).

***Synthesis of $Ni_3(HHTP)_2$ and $Co_3(HHTP)_2$***

$Ni(OAc)_2 \times 4H_2O$ (0.099 g, 0.4 mmol, 2 eq) and HHTP (0.065 g, 0.2 mmol, 1 eq) were dissolved in $H_2O$ (40 mL). The reaction mixture was sonicated for 20 minutes and then heated at 85 °C for 24 hours. The solid residue was washed with water, ethanol and acetone. The product was activated with ethanol (4×10 mL) and then dried in a vacuum oven at 50 °C overnight. $Ni_3(HHTP)_2$ was obtained as a black solid with a crystalline structure as follows from the powder X-ray diffraction (PXRD) analysis (Fig. S13a).

A similar procedure was used for the synthesis of $Co_3(HHTP)_2$. $Co(OAc)_2 \times 4H_2O$ (0.996 g, 0.4 mmol, 2 eq) and HHTP (0.065 g, 0.2 mmol, 1 eq) were dissolved in $H_2O$ (15 mL) and N-methylpyrrolidone (1.65 mL), the reaction mixture was sonicated for 20 minutes and then heated at 85 °C for 24 hours. The product was activated with ethanol (4×10 mL) and then dried in a vacuum oven at 50 °C overnight. $Co_3(HHTP)_2$ was obtained as a black solid with a crystalline structure as follows from the PXRD analysis (Fig S13b).

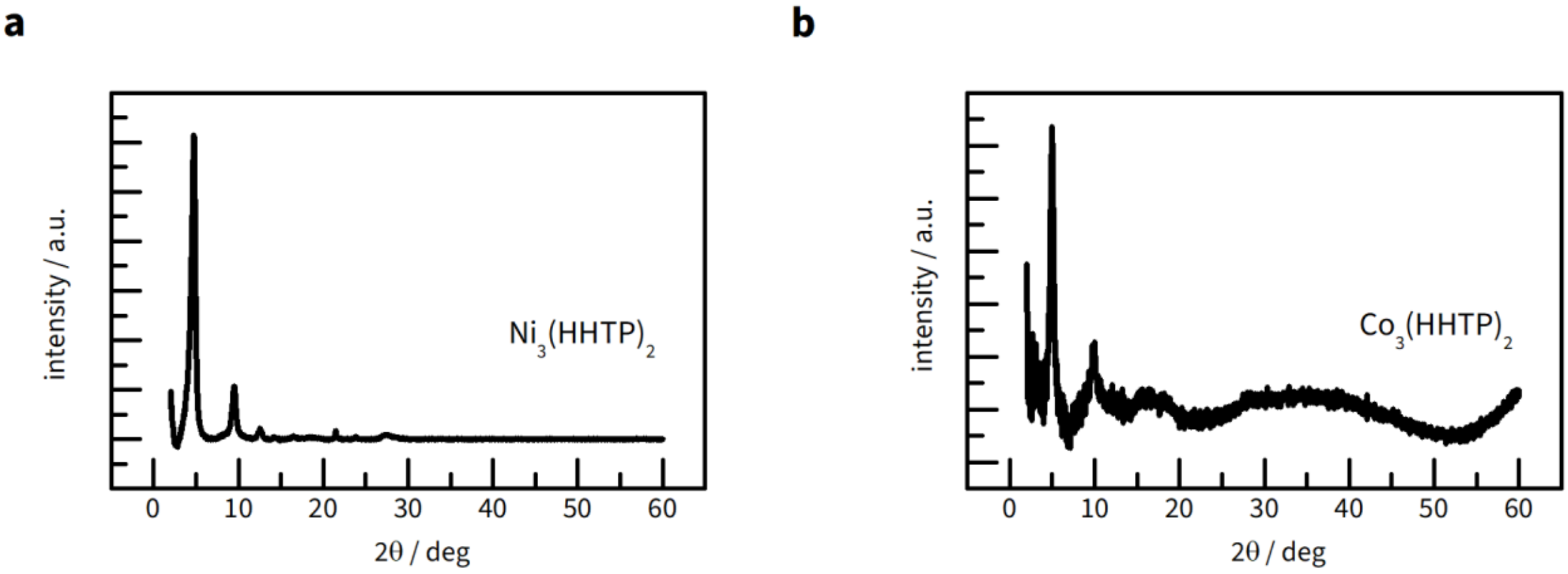

**Figure S14.** Powder X-ray diffraction (PXRD) patterns obtained for **(a)** $Ni_3(HHTP)_2$ and **(b)** $Co_3(HHTP)_2$.

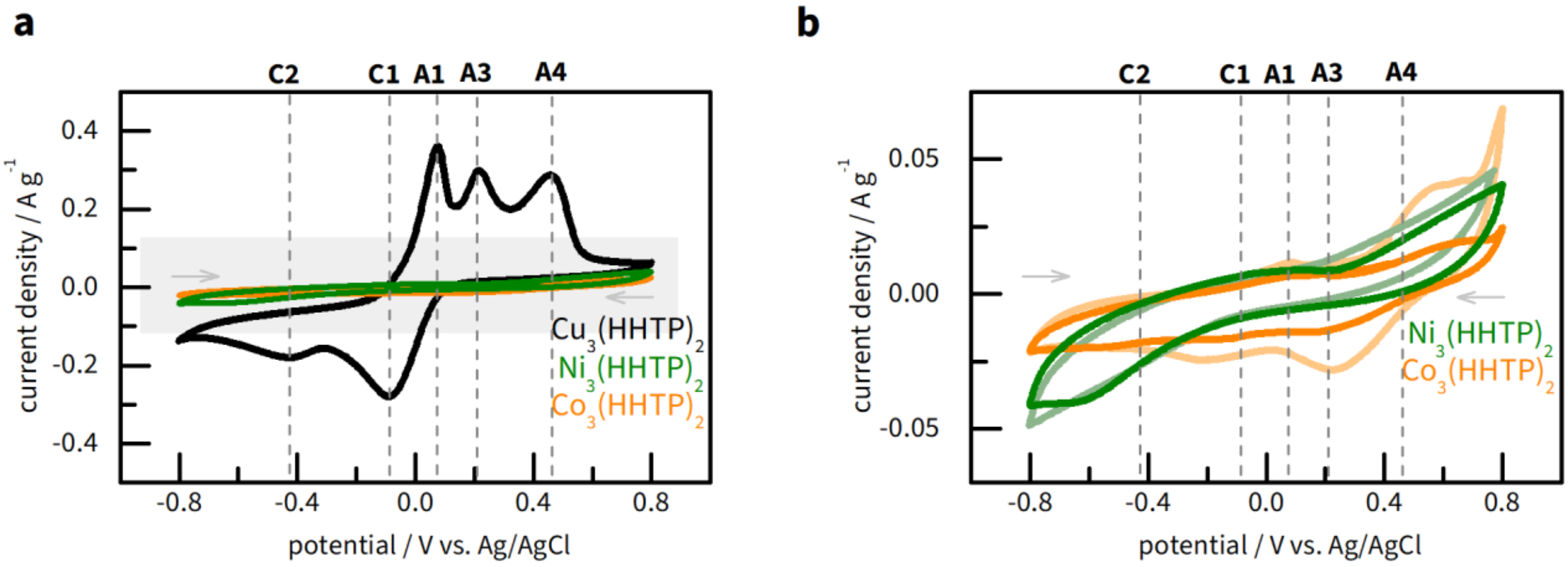

**Figure S15.** Cyclic voltammetry (CV) curves recorded at a scan rate of 10 mV s$^{-1}$ in 0.1 M $NaClO_4$. (a) CV curves in a $CO_2$-saturated electrolyte for different electrodes: black – GC-$Cu_3(HHTP)_2$, orange – GC-$Co_3(HHTP)_2$, and green – GC-$Ni_3(HHTP)_2$. (b) CV curves in an Ar-saturated electrolyte for different electrodes: light orange – GC-$Co_3(HHTP)_2$ and light green – GC-$Ni_3(HHTP)_2$; in a $CO_2$-saturated electrolyte for: dark orange – GC-$Co_3(HHTP)_2$ and dark green – GC-$Ni_3(HHTP)_2$.

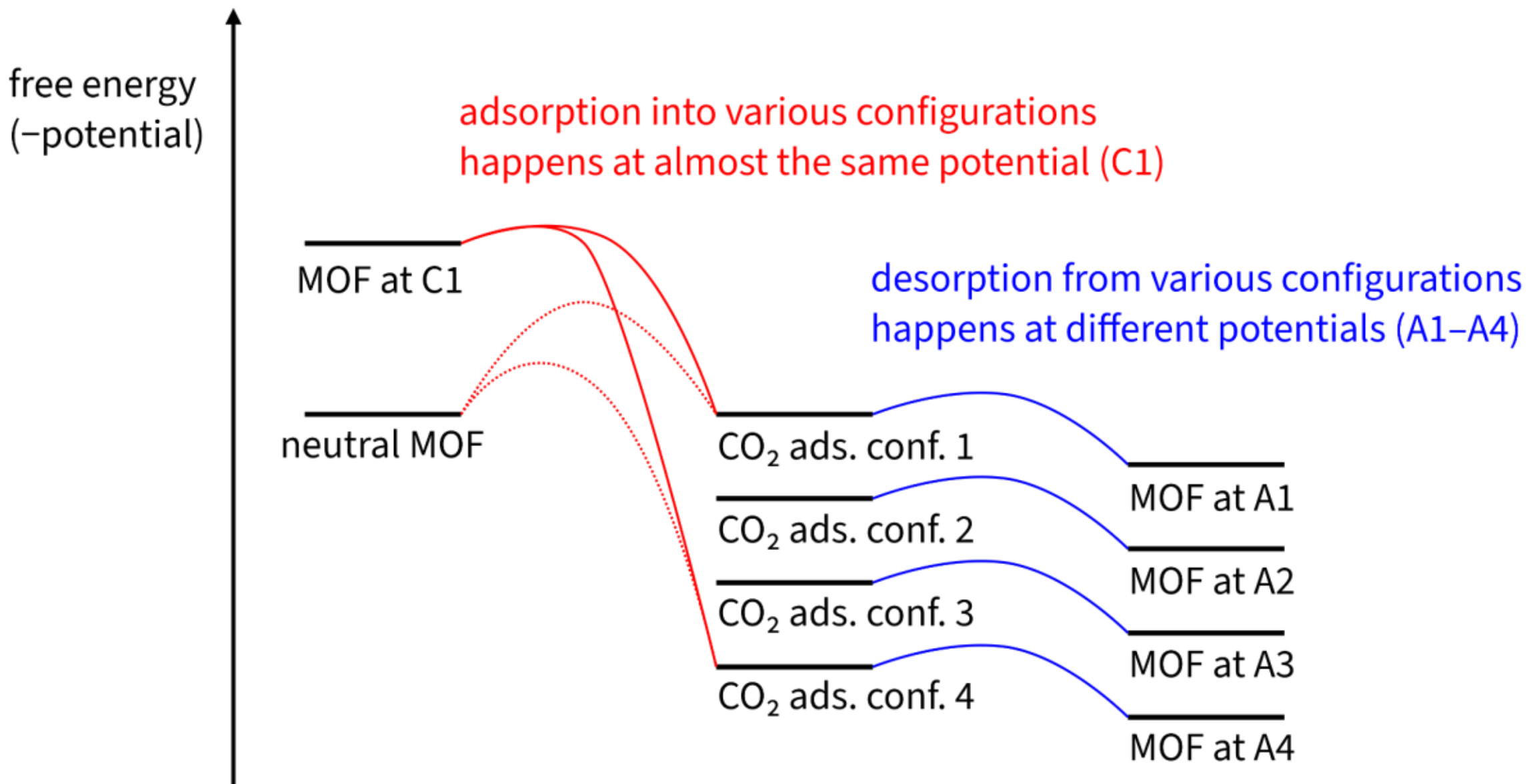

**Figure S16.** Free energy diagram suggesting that the distinct anodic peaks (A1–A4) are due to desorption from distinct configurations (1–4), in which adsorbed $CO_2$ has a variable number of neighbouring $CO_2$ molecules, i.e., different surroundings. We assume that the adsorption at cathodic peak (C1) is governed by the electron transfer from copper to the adsorbing $CO_2$, which is favoured for forming multiple configurations. The desorption from these configurations requires different potentials.

***Density Functional Theory calculations***

For all Density Functional Theory (DFT) calculations, open source ASE 3.23.0 and GPAW 24.1.0 packages were used,[19,20] with the RPBE exchange-correlation functional,[21] along with the D4 dispersion correction.[22] The core and valence electrons were described with the projector augmented-wave method. Standard 24.1.0 setups were used for all elements along double-zeta basis sets for obtaining the initial density, and Hubbard U of 10.4 eV for Cu.[23] Grid spacing of ~0.12 Å with a number of grid points divisible by 4 were used. Spin-polarisation was turned on, and the magnetic moment of ±1 was preset on $Cu^{2+}$ ions, with alternating signs for distinct layers. Brillouin-zone sampling was made with 4 k-points in the MOF plane and 2 k-points in perpendicular directions. The named parameters were tested for energy convergence. For all other parameters, default values were taken.

$CO_2$ adsorption was modelled in finite-difference mode using a solvated jellium model.[24] A constant counter charge of +1*e* per Cu atom was located in the jellium region, 3.5–6.5 Å above the Cu atom. An implicit water layer was put into a cavity between the jellium and MOF to solvate $CO_2$.[25] In the case of the three-dimensional model of $Cu_3(HHTP)_2$ crystal, only the terminal facet was studied for adsorption due to the limitation of the solvated jellium model, which is currently applicable only to flat electrochemical interfaces. To compare adsorption on the facet and in the pore, simplified two- and one-dimensional slab models were constructed (Fig. S2). Three adsorption sites were considered: above the Cu atom, above the O atom, and above an aromatic ring.

The three-dimensional model of $Cu_3(HHTP)_2$ crystal was created using *in silico* data from the EC-MOF database.[26] Two- and one-dimensional models, mimicking $Cu_3(HHTP)_2$, were created using the optimised parameters of the three-dimensional model (Fig. S2). Optimisation of atomic positions and unit cell parameters of pure MOF models was made in plane wave mode with a cut-off of 600 a.u. and the FrechetCellFilter,[1] until the residual forces of each atom become less than 0.1 eV $Å^{-1}$. Optimisation of atomic positions of MOF models with adsorbed $CO_2$ was run with the Broyden–Fletcher–Goldfarb–Shanno algorithm until the residual forces on each atom became less than 0.1 eV $Å^{-1}$. Enthalpy, entropy, and free energy values were evaluated at ideal gas and harmonic approximations through a vibrational analysis as implemented in the ASE thermochemistry module.